\documentclass[letterpaper, 10 pt, conference]{ieeeconf}  

\IEEEoverridecommandlockouts                              

\usepackage{graphics} 
\usepackage{epsfig} 
\usepackage{amsmath} 
\usepackage{amssymb}  
\usepackage{pgfplots}
\pgfplotsset{width=7cm,compat=newest}
\usepackage{subcaption}
\usepackage{units}

\usepackage[nolist]{acronym}

\title{\LARGE \bf Learning Super-resolved Depth from Active Gated Imaging
}

\author{Tobias Gruber$^{1,2}$, Mariia Kokhova$^{1,3}$, Werner Ritter$^{1}$, Norbert Haala$^{3}$ and Klaus Dietmayer$^{2}$%
\thanks{$^{1}$ The authors are with Daimler AG, RD/AFU, Wilhelm-Runge-Str. 11, 89081 Ulm, Germany, {\tt\scriptsize tobias.gruber@daimler.com}, {\tt\scriptsize mariia.kokhova@daimler.com}, {\tt\scriptsize werner.r.ritter@daimler.com}}
\thanks{$^{2}$ The authors are with the Institute of Measurement, Control and Microtechnology, Ulm University, Albert-Einstein-Allee 41, 89081 Ulm, Germany {\tt\scriptsize klaus.dietmayer@uni-ulm.de}}
\thanks{$^{3}$ The authors are with the Institute for Photogrammetry, University of Stuttgart, Geschwister-Scholl-Str. 24D, 70174 Stuttgart, Germany {\tt\scriptsize norbert.haala@ifp.uni-stuttgart.de}}
}

\begin{document}


\definecolor{dai_ligth_grey}{RGB}{230,230,230}
\definecolor{dai_ligth_grey20K}{RGB}{200,200,200}
\definecolor{dai_ligth_grey40K}{RGB}{158,158,158}
\definecolor{dai_ligth_grey60K}{RGB}{112,112,112}
\definecolor{dai_ligth_grey80K}{RGB}{68,68,68}

\definecolor{dai_petrol}{RGB}{0,103,127}
\definecolor{dai_petrol20K}{RGB}{0,86,106}
\definecolor{dai_petrol40K}{RGB}{0,67,85}
\definecolor{dai_petrol80}{RGB}{0,122,147}
\definecolor{dai_petrol60}{RGB}{80,151,171}
\definecolor{dai_petrol40}{RGB}{121,174,191}
\definecolor{dai_petrol20}{RGB}{166,202,216}

\definecolor{dai_deepred}{RGB}{113,24,12}
\definecolor{dai_deepred20K}{RGB}{90,19,10}
\definecolor{dai_deepred40K}{RGB}{68,14,7}

\definecolor{apfelgruen}{RGB}{140, 198, 62}
\definecolor{orange}{RGB}{244, 111, 33}
\definecolor{anthrazit}{RGB}{19, 31, 31}

\begin{acronym}
 \acro{CNN}{convolutional neural network}
 \acro{CMOS}{complementary metal-oxide semiconductor}
 \acro{EU}{European Union}
 \acro{DENSE}{aDverse wEather eNvironment Sensing systEm}
 \acro{FIR}{far infrared}
 \acro{NIR}{near infrared}
 \acro{SWIR}{short wave infrared}
 \acro{ADAS}{automotive drive assistance system}
 \acro{RMS}{root mean squared}
 \acro{ZNCC}{zero-mean normalized cross correlation}
 \acro{HDR}{high dynamic range}
 \acro{NN}{neural network}
 \acro{ECC}{error-correcting code}
 \acro{MLD}{maximum likelihood decoding}
 \acro{HDD}{hard decision decoding}
 \acro{SDD}{soft decision decoding}
 \acro{NND}{neural network decoding}
 \acro{ML}{maximum likelihood}
 \acro{GPU}{graphical processing unit}
 \acro{BP}{belief propagation}
 \acro{LDPC}{low density parity check}
 \acro{BER}{bit error rate}
 \acro{SNR}{signal-to-noise-ratio}
 \acro{ReLU}{rectified linear unit}
 \acro{BPSK}{binary phase shift keying}
 \acro{AWGN}{additive white Gaussian noise}
 \acro{MSE}{mean squared error}
 \acro{LLR}{log-likelihood ratio}
 \acro{MAP}{maximum a posteriori}
 \acro{NVE}{normalized validation error}
 \acro{BCE}{binary cross-entropy}
 \acro{BLER}{block error rate}
 \acro{IoT}{internet of things} 
 \acro{GDP}{gate delay profile}
 \acro{RIP}{range intensity profile}
 \acro{lidar}{light detecting and ranging}
 \acro{BWV}{BrightWayVision}
 \acro{MAE}{mean absolute error}
 \acro{NN}{neural network}
\end{acronym}


\maketitle
\thispagestyle{empty}
\pagestyle{empty}

\begin{abstract}
Environment perception for autonomous driving is doomed by the trade-off between range-accuracy and resolution: current sensors that deliver very precise depth information are usually restricted to low resolution because of technology or cost limitations.
In this work, we exploit depth information from an active gated imaging system based on cost-sensitive diode and CMOS technology.
Learning a mapping between pixel intensities of three gated slices and depth produces a super-resolved depth map image with respectable relative accuracy of 5\,\% in between \unit[25--80]{m}.
By design, depth information is perfectly aligned with pixel intensity values. 
\end{abstract}

\section{Introduction}

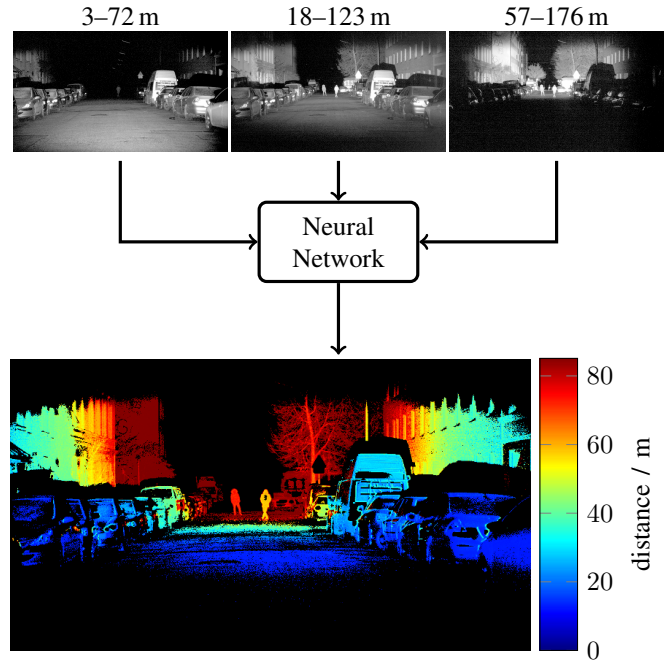
\begin{figure}
\centering
\begin{tikzpicture}
\tikzstyle{box} = [draw,very thick,rounded corners=.1cm,inner sep=5pt,minimum height=3em, text width=5em, align=center] 
\tikzstyle{circ} = [draw=green, very thick,circle,minimum size=1.5cm]

\node (slice1_txt) at (1.1,1) {\unit[3--72]{m}};

\node (slice2_txt) at (4,1) {\unit[18--123]{m}};

\node (slice3_txt) at (6.9,1) {\unit[57--176]{m}};

\node (slice1) at (1.1,0) {\includegraphics[width=0.33\columnwidth]{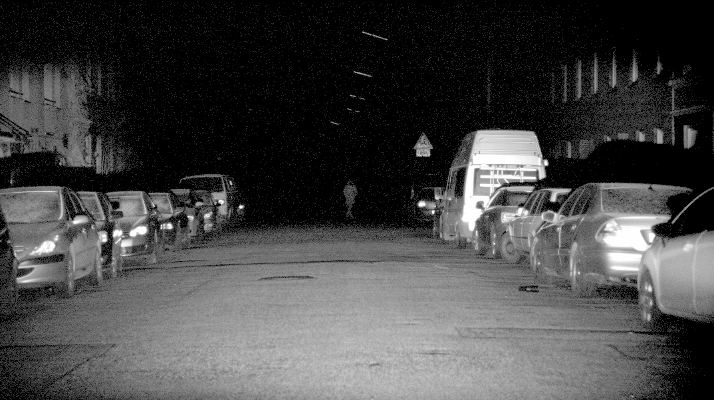}};

\node (slice2) at (4,0) {\includegraphics[width=0.33\columnwidth]{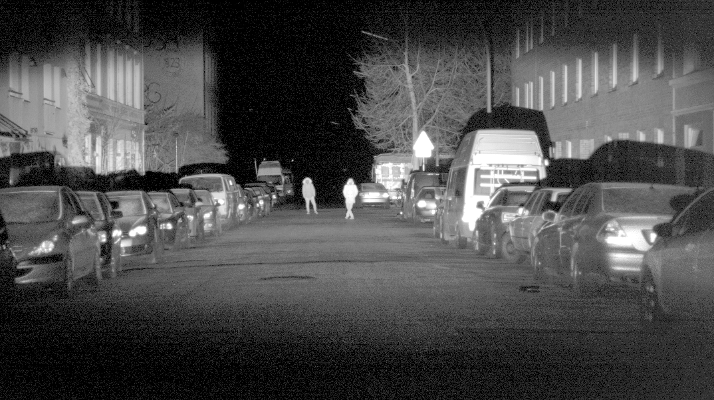}};

\node (slice3) at (6.9,0) {\includegraphics[width=0.33\columnwidth]{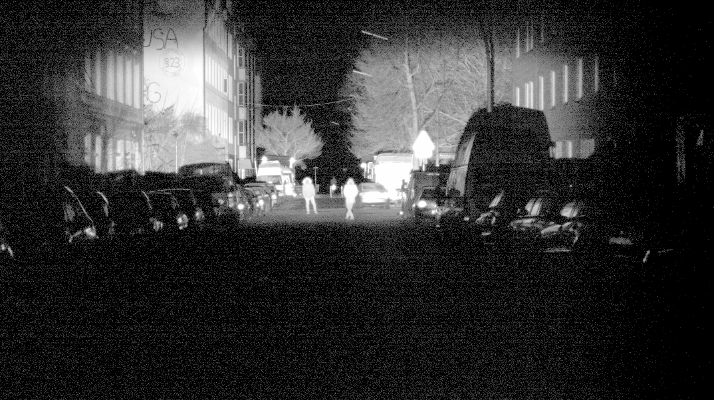}};

\node (depthmap) at (3.1,-5.5) {\includegraphics[width=0.8\columnwidth]{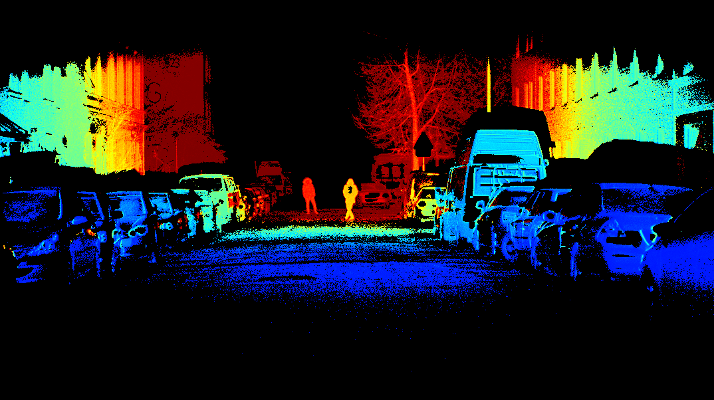}};

\node[box] (nn) at (4,-2) {Neural Network};

\draw[very thick, ->] (slice1) |- (nn);
\draw[very thick, ->] (slice2) -- (nn);
\draw[very thick, ->] (slice3) |- (nn);
\draw[very thick, ->] (nn) -- (4,-3.5);

\begin{axis}[
			hide axis,
			scale only axis,
			height=0.45\columnwidth,
			at={(depthmap.east)},
			yshift=19.5mm,
			xshift=-3mm,
			anchor=north,
			colormap/jet,
			colorbar,
			point meta min=0, 
			point meta max=85.117119, 
			colorbar style={
			ylabel=distance / \unit{m},
			},
		]
			\addplot [draw=none] coordinates {(1,1)};
		\end{axis}
\end{tikzpicture}
\caption{A neural network (NN) processes three slices of an active gated imaging system and yields a depth map.}
\label{fig:main_idea}
\end{figure}

Safe autonomous driving at level five requires perfect environment perception under any conditions, even in bad weather.
Active gated imaging is a promising technology that improves vision especially in bad weather as it removes backscatter by time-synchronizing illumination and exposure. 
It was shown in an image sensor benchmark \cite{Bijelic2018} that active gated imaging outperforms standard imaging in foggy conditions. 
In addition to stable contrast in most weather conditions, active gated imaging can provide depth information by suitable post-processing.
Several approaches exist for accessing this depth information.
However, these approaches are mostly restricted to certain laser pulse forms and delays.
In this work, we describe an approach to estimate depth for a completely free-shaped \ac{RIP} based on \acp{NN}. 
As depicted in Fig.~\ref{fig:main_idea}, we use three overlapping slices in order to obtain a depth map.
Range is estimated pixel-wise by finding a nonlinear function that maps the three intensity values of the slices to a range. 
We use the combination of a gated camera system and a \ac{lidar} sensor in order to obtain labeled data by simply driving in a natural environment.
It is shown that a very simple single hidden layer neural network with 40 nodes is able to estimate range with approximately 5\,\% relative accuracy for a distance range in between \unit[25--80]{m}.
Our depth estimation can be described as super-resolved since the gated camera has a resolution of 1280x720 pixels.

\subsection{Related Work}

\subsubsection{Active Gated Imaging}

First successful feasibility tests for active gated imaging were first reported in 1967 by Heckmann and Hodgson \cite{Heckman1967}. 
The purpose of this technique was to improve efficiency of underwater photography by avoiding backscattering.
Their system consisted of a light source that emitted short high-intensity bursts of light and a receiver that could be opened and closed very fast.
Further development of this technology was driven mainly by military research and led to many applications such as target identification \cite{Driggers2003}, night vision \cite{Inbar2008}, underwater imaging \cite{Fournier1993} and three-dimensional imaging \cite{Busck2004}. 
In comparison to laser scanning, the result of active gated imaging is a two-dimensional image of the scene.
However, by capturing multiple gated images with different delays, it is also possible to obtain range information.
The first approach to estimate distance was done in order to identify sea mines on the seabed \cite{Busck2004}.
By using cost sensitive components such as laser diodes and a \ac{CMOS} imager, Brightway Vision wants to establish such a gated system for use as a night vision system on the automotive market \cite{Inbar2008}.
In this work, such a system is extended with three-dimensional imaging and evolves into a \emph{range camera}.

\subsubsection{Time Slicing Method}

First attempts to extract depth from multiple gated images were made at the beginning of the century especially driven by defense research.
Busck and Heiselberg \cite{Busck2004} introduced the \emph{time-slicing method} for range estimation which samples the \ac{GDP} temporally with a single target by increasing the delay in steps and estimates the range by a weighted average \cite{Busck2005}. 
Similar work has been done by He and Seet \cite{He2004} where they sampled the \ac{GDP} spatially with multiple targets at certain distances.
Andersson presented in \cite{Andersson2006} two other depth reconstruction methods based on a sampled \ac{GDP}: least-squares parameter fitting and data feature positioning.
In recent years, Chua et al. developed a noise-weighted average range calculation in order to handle noise influence better \cite{Chua2016}.
However, sampling the \ac{GDP} requires recording many images of the same scene. 
This is no problem if the scene is stationary, but not suitable for real-time depth estimation in changing scenes.

\subsubsection{Range-intensity Correlation Method}

Depth accuracy of the time slicing method strongly depends on a huge number of very short pulse widths and scanning step sizes \cite{Busck2004, Busck2005, Andersson2006}.
The work of Laurenzis et al. \cite{Laurenzis2007} tries to overcome this resolution limit by exploiting the trapezoidal shape of the \ac{RIP}. 
This \emph{range-intensity correlation method} requires a series of images with overlapping \acp{RIP}.
In the overlapping region of a plateau and a rising or falling ramp, one can estimate the pixel-wise distance according to the relation of the intensities in both images by linear interpolation.
They improved the previous method for non-trapezoidal \acp{RIP} and find a transformation rule from intensities to estimated distance by fitting a multi-order ratio-polynomial model \cite{Laurenzis2009}.
The linear interpolation method for trapezoidal shapes was generalized to triangular shaped \acp{RIP} in \cite{Xinwei2013}.
For all of these range-intensity correlation methods perfect prior knowledge of the \acp{RIP} is required. 
Due to imperfections at laser pulse and gate shape modeling, e.g. rise and fall times, this is often very hard to obtain.
Moreover, methods in \cite{Laurenzis2007, Laurenzis2009, Xinwei2013} are limited to a rectangular modeling of the laser pulse and gate shape.

\subsubsection{Gain Modulation Method}

The gain modulation method, introduced in 2008 by Xiuda et al. \cite{Xiuda2008}, is a method independent from the laser pulse shapes.
However, for this method it is important that the laser pulse width is much smaller than the gate duration.
By using a gain-modulated and a gain-constant image, it is possible to recover depth. 
The modulation can be linear \cite{Xiuda2008} or exponential \cite{Jin2009}.

\subsubsection{Recent Range Reconstruction Methods}

In recent years, correlated double sampling was developed \cite{Gohler2017} that is able of capturing two images with different delays from a single laser pulse. 
This hardware adaption makes range reconstruction methods faster.
Due to the rising interest in \acp{CNN}, there is also a lot of movement in the area of depth map prediction from a single image, e.g. \cite{Eigen2014}. 
However, these methods are not comparable to our approach as they either use special hardware or RGB images.

\section{Getting Depth from Gated Images}

\acresetall

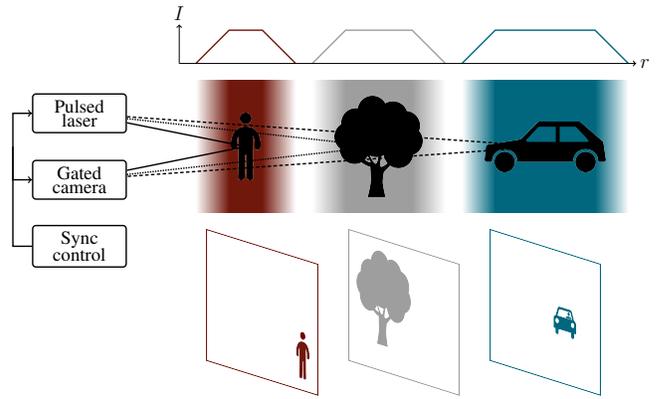
\begin{figure}
\centering
\resizebox{\columnwidth}{!}{
\begin{tikzpicture}
\tikzstyle{box} = [draw,very thick,rounded corners=.1cm,inner sep=5pt,minimum height=3em, text width=7em, align=center] 
\tikzstyle{circ} = [draw=green, very thick,circle,minimum size=1.5cm]

\node[box,draw=black] (laser) at (0,1) {\LARGE Pulsed laser};

\node[box,draw=black] (cam) at (0,-1) {\LARGE Gated camera};

\node[box,draw=black] (sync) at (0,-3) {\LARGE Sync control};

\coordinate (v0) at (-2,0) {};
\draw [very thick]  (sync) -| (v0);
\draw [very thick, ->]  (v0) |- (laser);
\draw [very thick, ->]  (v0) |- (cam);

\shade[left color=white,right color=dai_deepred] (3.5,-2) rectangle (4.5,2);
\fill[dai_deepred] (4.5,-2) rectangle (5.5,2);
\shade[left color=dai_deepred,right color=white] (5.5,-2) rectangle (6.5,2);

\shade[left color=white,right color=dai_ligth_grey40K] (7,-2) rectangle (8,2);
\fill[dai_ligth_grey40K] (8,-2) rectangle (10,2);
\shade[left color=dai_ligth_grey40K, right color=white] (10,-2) rectangle (11,2);

\shade[left color=white,right color=dai_petrol] (11.5,-2) rectangle (12.5,2);
\fill[dai_petrol] (12.5,-2) rectangle (15.5,2);
\shade[left color=dai_petrol,right color=white] (15.5,-2) rectangle (16.5,2);

\node (pedestrian) at (5,0) {\includegraphics[height=2cm]{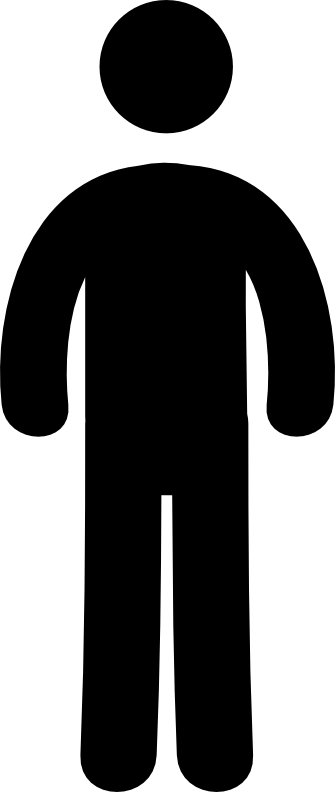}};
\coordinate (pedestrian_coord) at (pedestrian);
\draw [very thick]  (laser) -- (pedestrian_coord);
\draw [very thick]  (pedestrian_coord) -- (cam);

\node (tree) at (9,0) {\includegraphics[height=3cm]{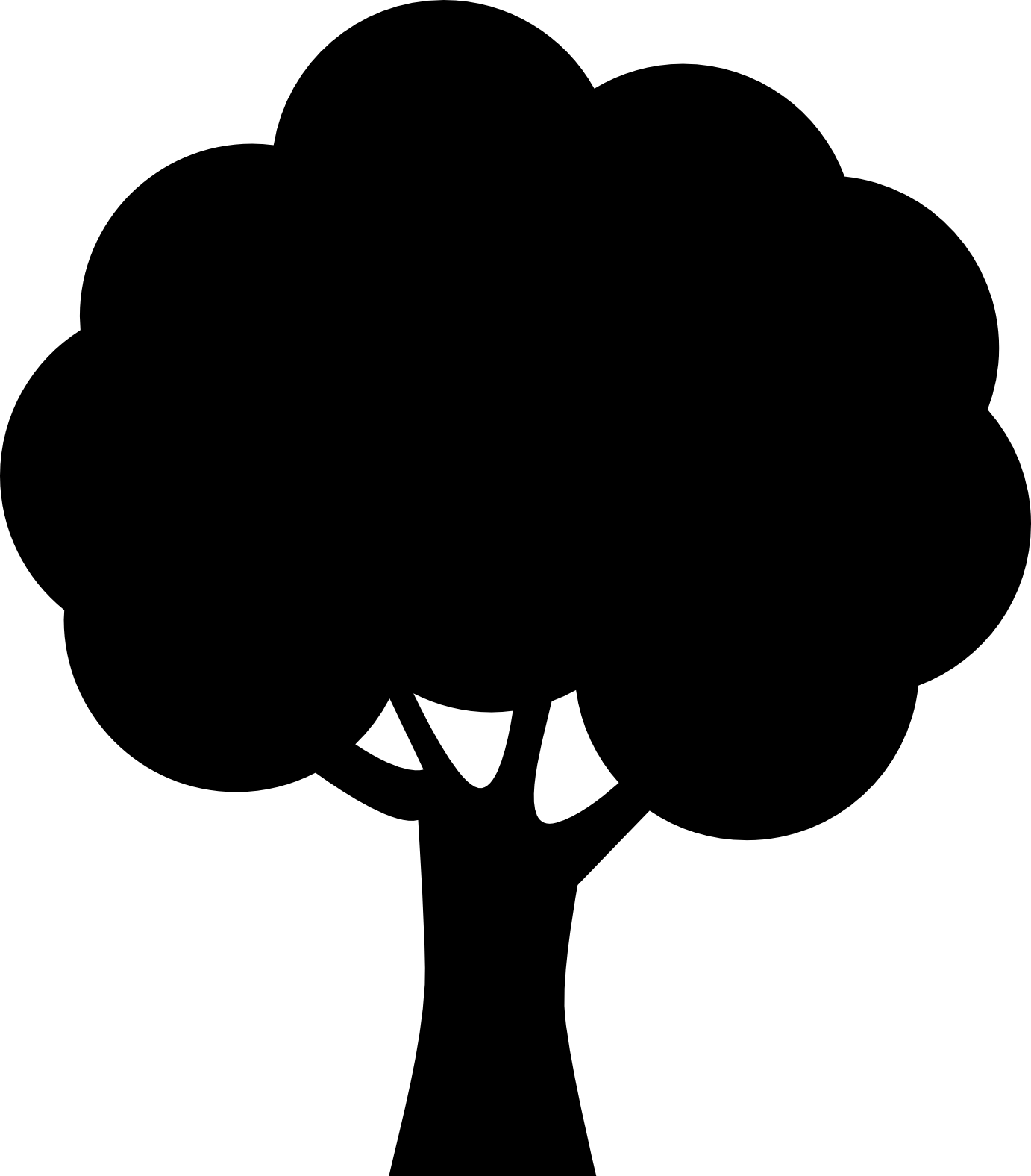}}; 
\coordinate (tree_coord) at (tree);
\draw [very thick, densely dotted]  (laser) -- (tree_coord);
\draw [very thick, densely dotted]  (tree_coord) -- (cam);

\node (car) at (14,0) {\includegraphics[height=1.5cm]{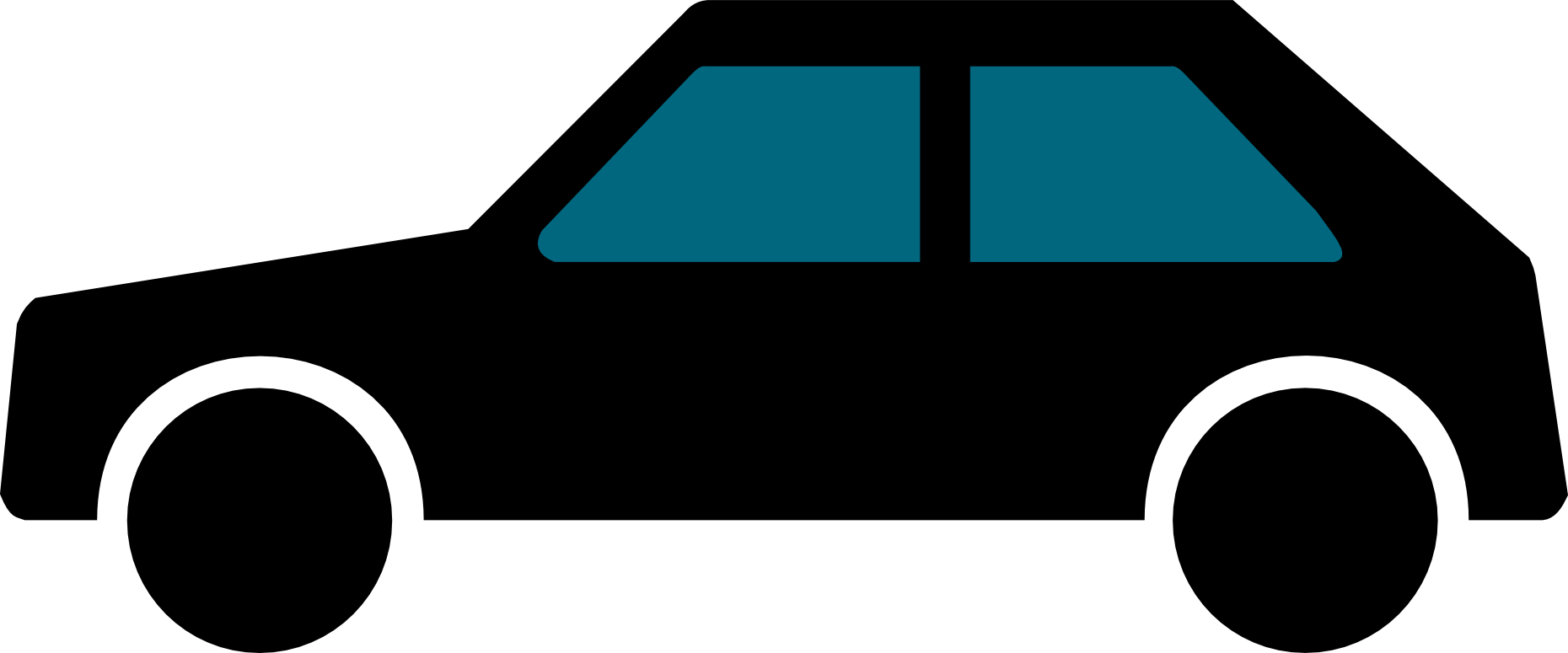}}; 
\coordinate (car_coord) at (car);
\draw [very thick, densely dashed]  (laser) -- (car_coord);
\draw [very thick, densely dashed]  (car_coord) -- (cam);

\draw [very thick, dai_deepred]  (3.5,2.5) -- (4.5,3.5) -- (5.5,3.5) -- (6.5,2.5);
\draw [very thick, dai_ligth_grey40K]  (7,2.5) -- (8,3.5) -- (10,3.5) -- (11,2.5);
\draw [very thick, dai_petrol]  (11.5,2.5) -- (12.5,3.5) -- (15.5,3.5) -- (16.5,2.5);

\node (r) at (17,2.5) {\LARGE $r$};
\coordinate (origin) at (3,2.5);
\node (I) at (3,4) {\LARGE $I$};
\draw [thick, ->]  (origin) -- (r);
\draw [thick, ->]  (origin) -- (I);

\node (slice_tree) at (5.5,-5) {\includegraphics[height=5cm]{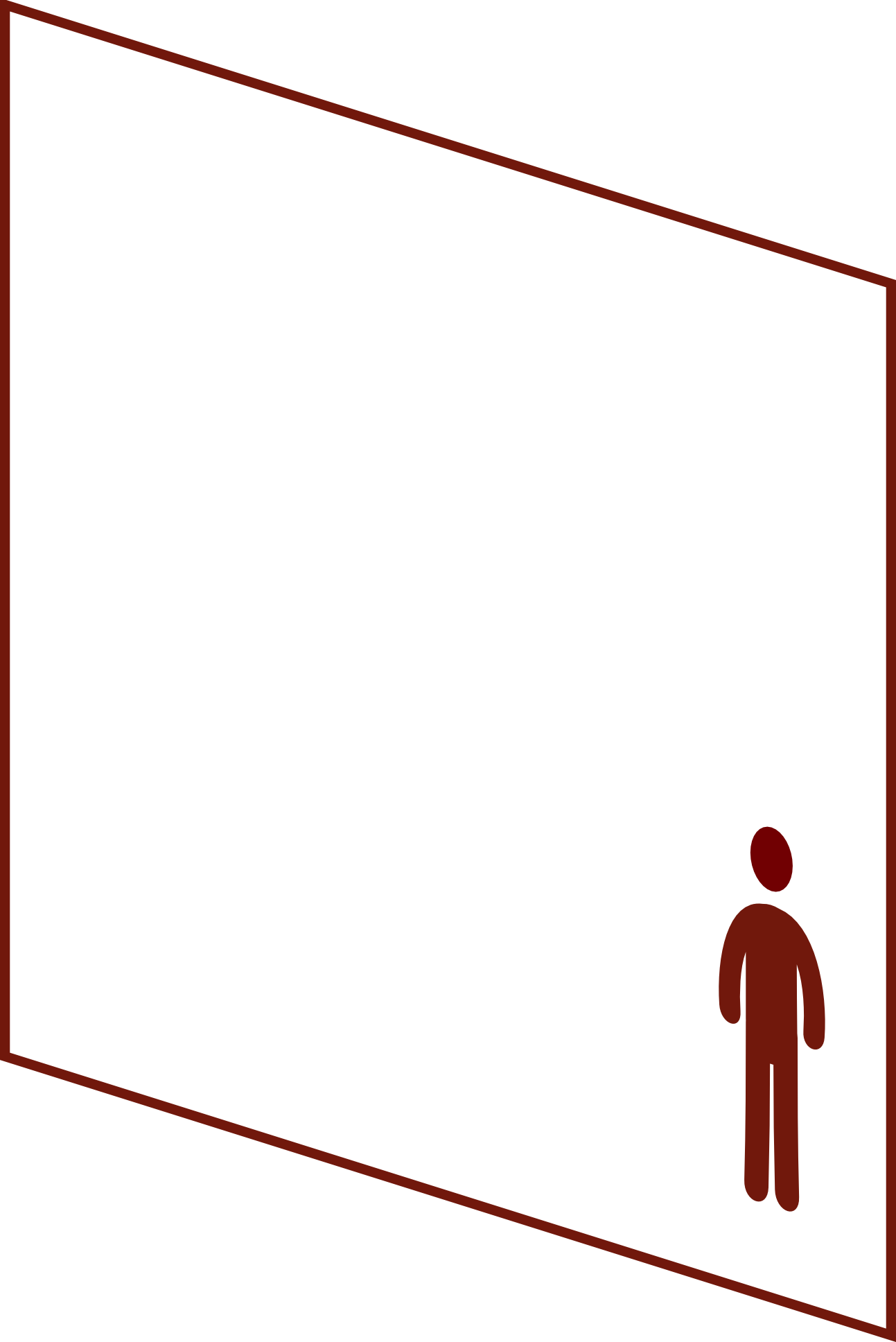}}; 

\node (slice_tree) at (9.75,-5) {\includegraphics[height=5cm]{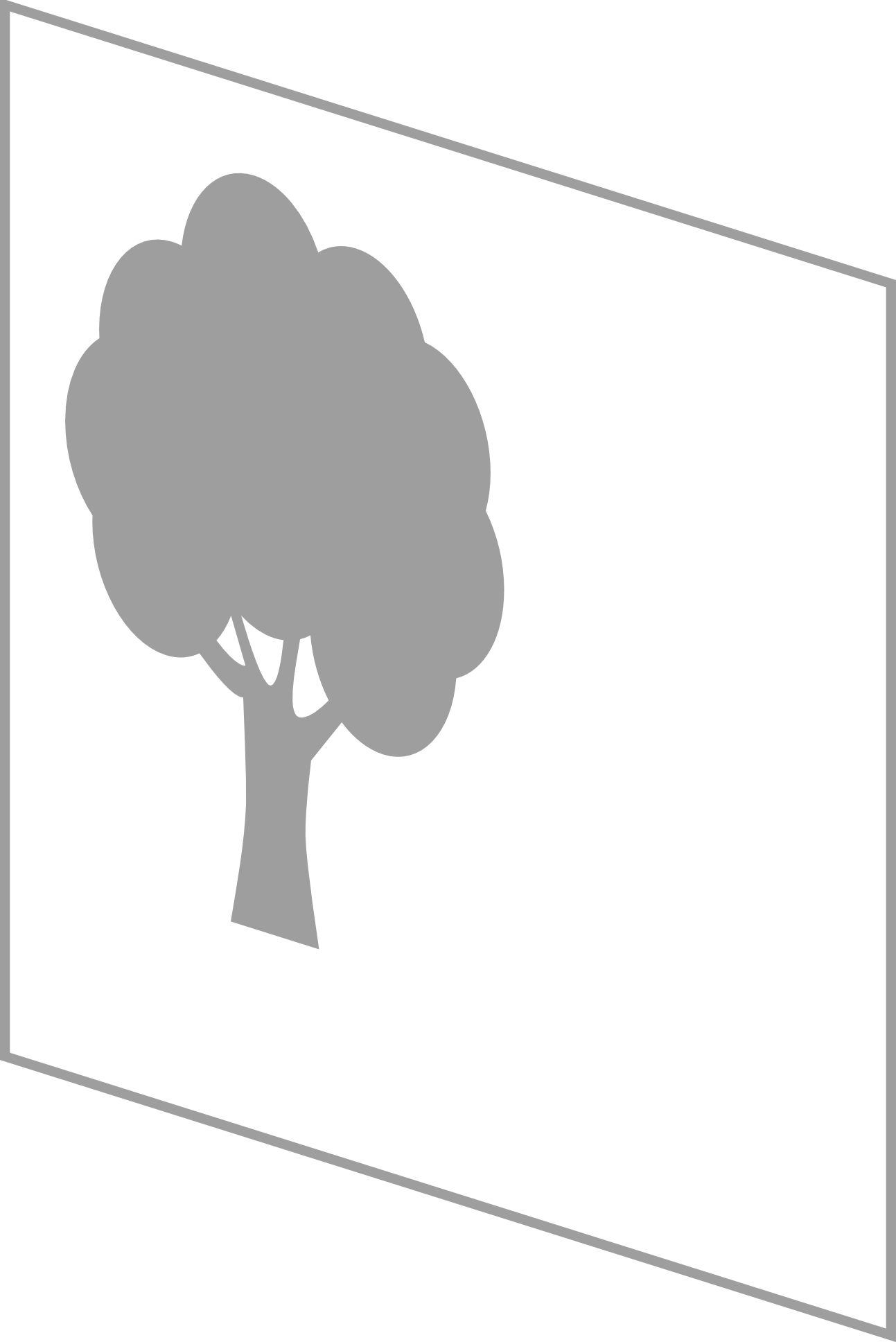}}; 

\node (slice_car) at (14,-5) {\includegraphics[height=5cm]{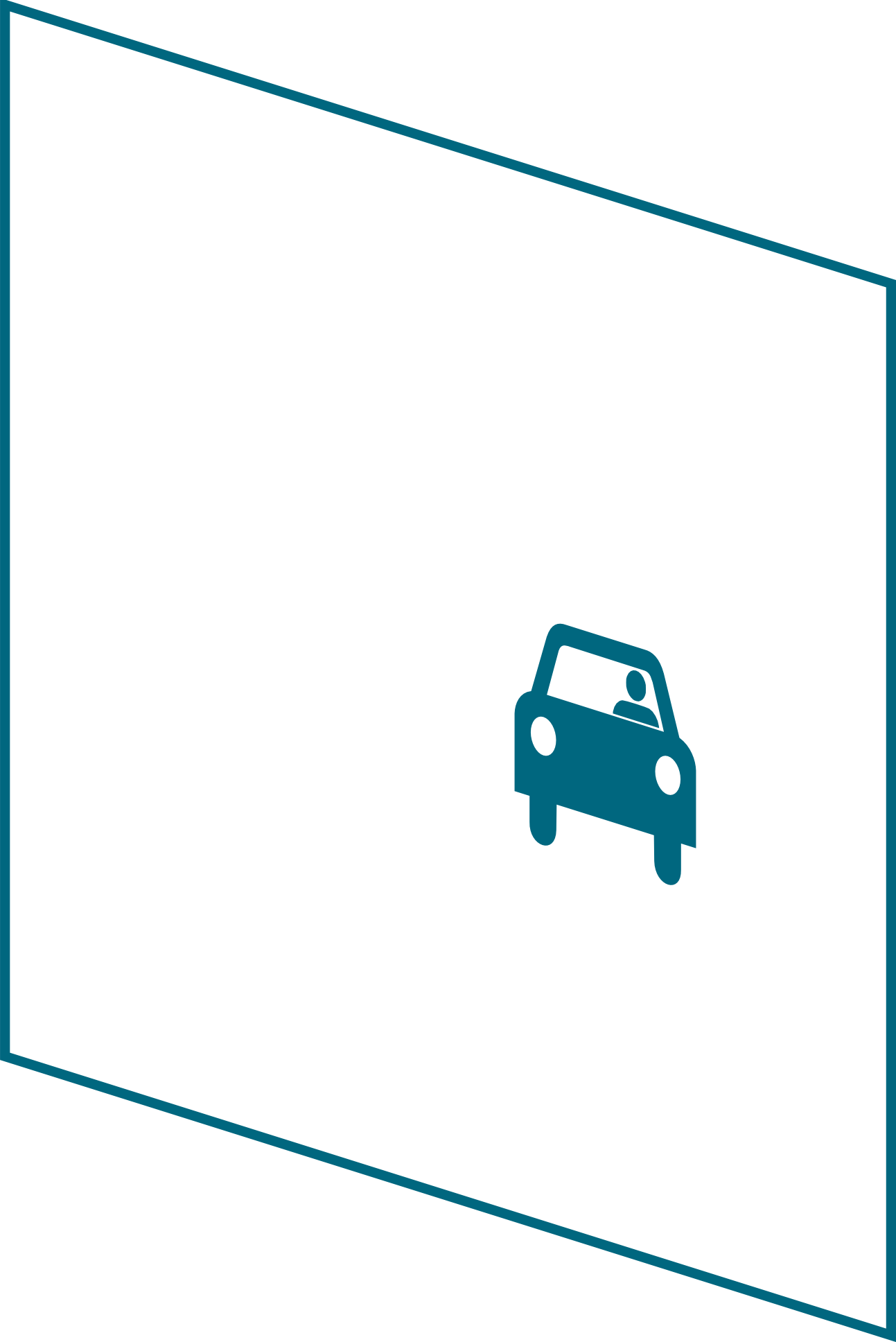}}; 
 
\end{tikzpicture}
}
\caption{Description of the active gated imaging system.}
\label{fig:gated_system}
\end{figure}

Active gated imaging is based upon a sensitive image sensor and its own illumination source as depicted in Fig.~\ref{fig:gated_system}. 
By time-synchronizing the image sensor and its illumination, it is possible to capture an image at a certain range, referred to as \emph{slice} in the following.
The scene is illuminated with a very short laser pulse and the camera gate opens after a designated delay $t_0$ in order to receive only photons from a certain distance $r$.
This helps to select the reflected light from objects and block back-scattered photons from clouds, fog or raindrops.
In order to improve the \ac{SNR} on the imager,  multiple number of pulses are usually collected on the chip \cite{Busck2004}.
Active gated imaging delivers images with high contrast even in extreme scattering conditions as in fog or underwater. 
However, active illumination means dealing with eye-safety restrictions and interference with similar systems.

\begin{figure}
	
	\begin{subfigure}[t]{\columnwidth}
		\subcaption{Laser pulse with $t_L=\unit[100]{ns}$.}
		\vspace{-2mm}
		\begin{tikzpicture}
				\begin{axis}
					[
					xlabel=time $t$ / \unit{ns},
					ylabel=power,
					ytick=\empty,
					yticklabels={,,},
					xmin=-100, xmax=600,
					width=\textwidth,
					height=0.12\textheight,
					]
				
				\addplot+ 	[
							very thick,
							solid, 
							dai_deepred,
							mark=none
							]
				table[x index=0,y index=1,col sep=space]{data/laser_pulse.data};
								
				\end{axis}
			\end{tikzpicture}
			
			\label{fig:laser_pulse}
	\end{subfigure}\vspace{3mm}
	\begin{subfigure}[t]{\columnwidth}
		\subcaption{Gate with $t_G=\unit[200]{ns}$.}
		\vspace{-2mm}
			\begin{tikzpicture}
					\begin{axis}
						[
						xlabel=time $t$ / \unit{ns},
						ylabel=gain,
						ytick=\empty,
						yticklabels={,,},
						xmin=-100, xmax=600,
						width=\textwidth,
						height=0.12\textheight,
						]
					
					\addplot+ 	[
								very thick,
								solid, 
								dai_deepred,
								mark=none
								]
					table[x index=0,y index=1,col sep=space]{data/gate.data};					
					
					\end{axis}
				\end{tikzpicture}
				
				\label{fig:gate}
		\end{subfigure}\vspace{3mm}
		\begin{subfigure}[t]{\columnwidth}
		\subcaption{Gate delay profile for $r=\unit[50]{m}$.}
		\vspace{-2mm}
		\begin{tikzpicture}
				\begin{axis}
					[
					xlabel=delay $t_0$ / \unit{ns},
					ylabel=$I_\text{GDP} \left( t_0 \right)$,
					ytick=\empty,
					yticklabels={,,},
					scaled y ticks = false,
					xmin=0, xmax=800,
					width=\textwidth,
					height=0.12\textheight,
					]
				
				\addplot+ 	[
							very thick,
							solid, 
							dai_deepred,
							mark=none
							]
				table[x index=0,y index=1,col sep=space]{data/gate_delay_profile.data};
				
				\end{axis}
			\end{tikzpicture}
			
			\label{fig:gate_delay_profile}
	\end{subfigure}\vspace{3mm}
	\begin{subfigure}[t]{\columnwidth}
		\subcaption{Range intensity profile for $t_0=\unit[100]{ns}$.}
		\vspace{-2mm}
		\begin{tikzpicture}
				\begin{axis}
					[
					xlabel=distance $r$ / \unit{m},
					ylabel=$I_\text{RIP} \left( r \right)$,
					ytick=\empty,
					yticklabels={,,},
					scaled y ticks = false,
					xmin=0, xmax=80,
					width=\textwidth,
					height=0.12\textheight,
					]
				
				\addlegendimage{very thick, solid, dai_deepred}
				
				\addplot+ 	[
							very thick,
							solid, 
							dai_deepred,
							mark=none
							]
				table[x index=0,y index=1,col sep=space]{data/range_intensity_profile.data};

				\end{axis}
			\end{tikzpicture}			
			\label{fig:range_intensity_profile}
	\end{subfigure}\vspace{3mm}
	\begin{subfigure}[t]{\columnwidth}
		\subcaption{\ac{RIP} with irradiance for $t_0=\unit[100]{ns}$.}
		\vspace{-2mm}
		\begin{tikzpicture}
				\begin{axis}
					[
					xlabel=distance $r$ / \unit{m},
					ylabel=$\hat{I}_\text{RIP} \left( r \right)$,
					ytick=\empty,
					yticklabels={,,},
					xmin=0, xmax=80,
					width=\textwidth,
					height=0.12\textheight,
					]
				
				\addlegendimage{very thick, solid, dai_deepred}
				
				\addplot+ 	[
							very thick,
							solid, 
							dai_deepred,
							mark=none
							]
				table[x index=0,y index=1,col sep=space]{data/range_intensity_profile_irr.data};				
				
				\end{axis}
			\end{tikzpicture}
			
			\label{fig:range_intensity_profile_irr}
	\end{subfigure}
	\caption{Example simulation of \acf{GDP} and \acf{RIP}. Y-axis can be arbitrary scaled.}
	\label{fig:pulses}
\end{figure}
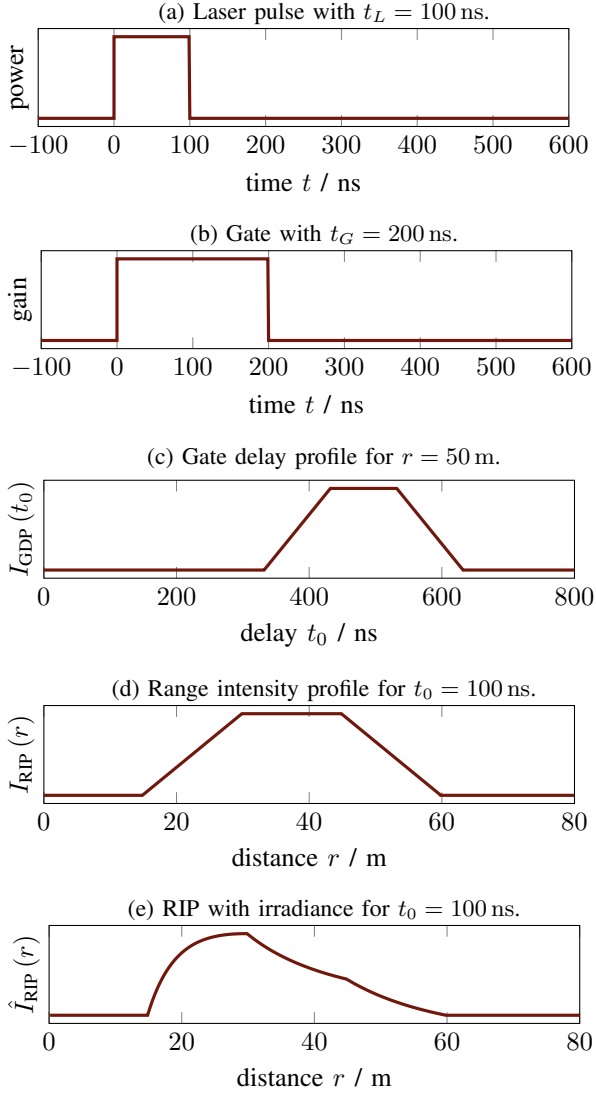

For the sake of simplicity, suppose that the shape of the laser pulse $p\left( t-2r/c_0 \right)$ and the detector gain $g\left( t-t_0 \right)$ are rectangular with length $t_L$ and $t_G$ respectively (see Fig.~\ref{fig:laser_pulse} and \ref{fig:gate}), where $c_0$ denotes the speed of light.
Following \cite{Andersson2006}, the pixel value $I\left( t_0, r \right)$ is proportional to the convolution of $p$ and $g$, thus 
\begin{align}
I\left( t_0, r \right) \propto \int\limits_{-\infty}^{\infty} g\left( t - t_0 \right) p\left( t - \frac{2r}{c_0} \right) \textrm{d}t .
\end{align}

\subsection{Gate Delay Profile (GDP)}

The \ac{GDP} $I_\text{GDP} \left( t_0 \right)$ describes the pixel intensity of an object at a certain distance $r$ if the delay $t_0$ is varied. 
If $t_L = t_G$ the shape of $I_\text{GDP} \left( t_0 \right)$ is triangular, otherwise trapezoidal (see Fig.~\ref{fig:gate_delay_profile}).
As introduced in \cite{Busck2005}, depth can be estimated from the \ac{GDP} by \emph{time-slicing}, that means sampling the \ac{GDP} by increasing the camera delay $t_0$ in $n$ steps with $\Delta t \ll t_L, t_G$ and estimate the depth by a conventional weighted average method. The average two-way travel time $\hat{t}$ is obtained by
\begin{align}
\hat{t} = \frac{\sum\limits_{i=1}^{n} I_i t_i}{\sum\limits_{i=1}^{n} I_i}
\end{align} 
where $I_i$ is the pixel intensity and $t_i = t_0 + i \Delta t$ the delay of slice $i$.
and the depth of the pixel $\hat{r}_\text{ts}$ can be reconstructed by
\begin{align}
\hat{r}_\text{ts} = \frac{c_0 \hat{t}}{2} .
\end{align}
However, the time-slicing method assumes a Gaussian laser pulse shape.
Literature shows that usually 10 to 100 slices are used for sampling a small range at close distance, resulting in an accuracy of \unit[1]{mm} \cite{Busck2004, Busck2005, Andersson2006}. 
Increasing the range of depth estimation would require many more slices or else result in lower accuracy.
Therefore, high accuracy and large depth of field cannot be realized simultaneously.

\subsection{Range Intensity Profile (RIP)}

In contrast to the \ac{GDP}, the \ac{RIP} $I_\text{RIP} \left( r \right)$ characterizes the pixel intensity for a fixed slice with delay $t_0$ with respect to distance $r$. 
Similar to the \ac{GDP}, the shape of the \ac{RIP} is basically trapezoidal for $t_L \neq t_G$, see Fig.~\ref{fig:range_intensity_profile}.
However, the pixel value of the \ac{RIP} additionally depends on the laser pulse irradiance $J\left(r\right)$, the reflectance $\alpha$ of the target and the influence of the atmosphere $\beta\left( r \right)$. 
Therefore, one can write $I_\text{RIP} \left( r \right)$ as
\begin{align}
\hat{I}_\text{RIP} \left( r \right) \propto \kappa \left( r \right) \int\limits_{-\infty}^{\infty} g\left( t - t_0 \right) p\left( t - \frac{2r}{c_0} \right) \textrm{d}t 
\end{align}
where $\kappa\left( r \right)$ is a distance dependent factor given by
\begin{align}
\kappa\left( r \right) = J\left( r \right) \alpha \beta \left( r \right) = \frac{1}{r^2} \alpha \beta \left( r \right) .
\end{align}
An example for $I_\text{RIP} \left( r \right)$ is given in Fig.~\ref{fig:range_intensity_profile_irr}.

The idea of \emph{range-intensity correlation} is to exploit the spatial correlation of overlapped gated images. 
For an object at a certain distance $r_0$ the factor $\kappa \left( r_0 \right)$ is constant and can be removed by considering only the intensity ratios.

\begin{figure}
		\begin{tikzpicture}
				\begin{axis}
					[
					xlabel=distance $r$ / \unit{m},
					ylabel=$I_\text{RIP} \left( r \right)$,
					xmin=10,
					xmax=90,
					xtick={15, 30, 45, 60, 75},
					xticklabels={$t_0$,$t_0 + t_L$, $t_0 + 2t_L$, $t_0 + 3t_L$, $t_0 + 4t_L$},
					xticklabel style = {font=\footnotesize},
					ytick=\empty,
					yticklabels={,,},
					legend style={
						cells={anchor=west},
						legend pos=north east,
						font=\scriptsize
					},
					legend entries={slice $i$, slice $i+1$},
					width=1.1\columnwidth,
					height=0.14\textheight,
					]
				
				\addlegendimage{very thick, solid, dai_deepred}
				\addlegendimage{very thick, solid, dai_petrol}
				
				\addplot+ 	[
							very thick,
							solid, 
							dai_deepred,
							mark=none
							]
				table[x index=0,y index=1,col sep=space]{data/rip_ric_example_slice1.data};
				
				\addplot+ 	[
							very thick,
							solid, 
							dai_petrol,
							mark=none
							]
				table[x index=0,y index=1,col sep=space]{data/rip_ric_example_slice2.data};

				\end{axis}
			\end{tikzpicture}
			\caption{Example \acfp{RIP} for two similar slices with $t_G = 2t_L$.}
			\label{fig:range_intensity_correlation}
	\end{figure}
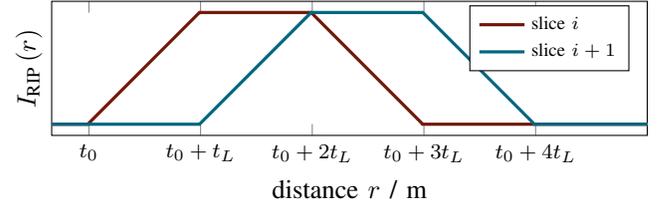

Suppose that for two slices $i$ and $i+1$ with gate duration $t_G = 2 t_L$ and delay $t_i=t_0$ and $t_{i+1} = t_0 + t_L$, then $I_i \left( r \right)$ and $I_{i+1} \left( r \right)$ are trapezoidal and overlapping, see Fig.~\ref{fig:range_intensity_correlation}.
According to \cite{Laurenzis2007}, depth information for distance where slice $i$ is a plateau and slice $i+1$ is rising, depth can be reconstructed by
\begin{align}
\hat{r}_\text{corr,trapez} = \frac{c_0}{2} \left( t_0 + t_L + \frac{I_i}{I_{i+1}} t_L \right) . \label{eq:corr_trapez}
\end{align}
For triangular pulse shapes, a plateau is defined by $I_{i} + I_{i+1}$ and depth for distances where both slices are rising is estimated by
\begin{align}
\hat{r}_\text{corr,triangle} = \frac{c_0}{2} \left( t_0 + t_L + \frac{I_i}{I_i + I_{i+1}} t_L \right) \label{eq:corr_triangle} .
\end{align}
These equations can be adapted to any region where two \acp{RIP} overlap.

This method works especially well for gated images with special pulse shapes, i.e. trapezoidal \cite{Laurenzis2007} or triangular \cite{Xinwei2013}, and adapted delays to generate these overlapping regions.
For a depth scene of \unit[650--1250]{m}, an accuracy of about \unit[30]{m} was achieved in \cite{Laurenzis2009}.

In order to get rid of the assumption of rectangular pulses, the \emph{gain modulation method} was proposed in \cite{Xiuda2008} and \cite{Jin2009}.
Two gated images are required for this method, i.e. a gain modulated image and a gain constant image. 
Depth can then be estimated, similar to range-intensity correlation, by the relation of both intensities.
In \cite{Xiuda2008}, depth accuracies around \unit[1]{m} are achieved for a range from \unit[800--1100]{m}.
However, this method requires special imagers that are capable of providing modulated gain.

\subsection{Why are these approaches not suitable?}

While time slicing is more appropriate for high accuracy and small ranges, range-intensity correlation and gain modulation have benefits for larger ranges at the expense of lower accuracy.
Gain modulation requires only two slices whereas range-intensity correlation requires more slices for long ranges and high accuracy. 
All methods prescribe conditions on certain pulse shapes, gate shapes and delays.

The idea of active gated imaging allows adapting the gating parameters so that the image offers maximal contrast at every distance.
For example, for larger distances longer exposure times helps to capture more photons.
Moreover, additional laser pulses at greater distances improve the \ac{SNR}.
If gating parameters are set up in order to improve the image quality, then current approaches have problems because their requirements are not satisfied anymore.

This work focuses on the automotive application of depth estimation from gated images.
Hence, we apply the gating parameters as provided in Table~\ref{tab:gating_parameters}, resulting in three \acp{RIP} as depicted in Fig.~\ref{fig:bwv_rip}.
\begin{figure}
		\begin{tikzpicture}
				\begin{axis}
					[
					xlabel=distance $r$ / \unit{m},
					ylabel=$I_\text{RIP} \left( r \right)$,
					ytick=\empty,
					yticklabels={,,},
					xmin=0, xmax=200,
					legend style={
						cells={anchor=west},
						legend pos=north east,
						font=\scriptsize
					},
					legend entries={slice1, slice2, slice3},
					width=1.05\columnwidth,
					height=0.17\textheight,
					]
				
				\addlegendimage{very thick, solid, dai_deepred}
				\addlegendimage{very thick, solid, dai_petrol}
				\addlegendimage{very thick, solid, dai_ligth_grey40K}
				
				\addplot+ 	[
							very thick,
							solid, 
							dai_ligth_grey40K,
							mark=none
							]
				table[x index=0,y index=3,col sep=space]{data/grayvalues.data};
				
				\addplot+ 	[
							very thick,
							solid, 
							dai_petrol,
							mark=none
							]
				table[x index=0,y index=2,col sep=space]{data/grayvalues.data};
				
				\addplot+ 	[
							very thick,
							solid, 
							dai_deepred,
							mark=none
							]
				table[x index=0,y index=1,col sep=space]{data/grayvalues.data};
				
				\end{axis}
			\end{tikzpicture}
			\caption{Range intensity profiles (RIPs) of our gated parameters as described in Table~\ref{tab:gating_parameters}.}
			\label{fig:bwv_rip}
\end{figure}
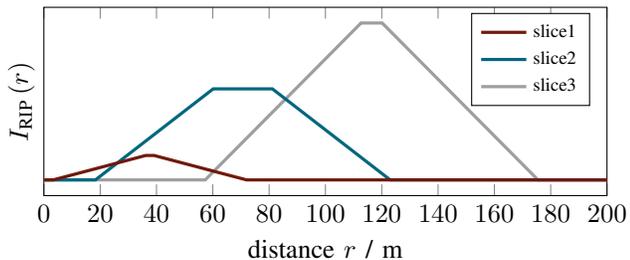
\begin{table}
	\centering
	\caption{Gating parameters.}
	\begin{tabular}{lrrr}
	& \textbf{slice1} & \textbf{slice2} & \textbf{slice3} \\ \hline
	pulses & 202 & 591 & 770 \\ 
	$t_L$ / \unit{ns} & 240 & 280 & 370 \\ 
	$t_G$ / \unit{ns} & 220 & 420 & 420 \\ 
	$t_0$ / \unit{ns} & 20 & 120 & 380 \\ 
	\end{tabular}
	\label{tab:gating_parameters}
\end{table}
In order to allow these flexible gated settings which can adapt to the appearance of the scene, new ideas have to be found for recovering depth from slices. 
In the following chapter, we propose a method to estimate depth for free-modeled gated profiles without any restrictions.

\subsection{Baseline Approach}

As a baseline our approach has to compete with, we implemented a range-intensity correlation algorithm for our special gating parameters: 
the distance range is divided into 9 sections where the behavior of the slices differ (rising, plateau, falling) and range is estimated according to Equations~\ref{eq:corr_trapez} and \ref{eq:corr_triangle}. 
If distance can be estimated in two ways, e.g. in between \unit[57--72]{m} as depicted in Fig.~\ref{fig:bwv_rip}, the distance estimate is given by the mean.
This approach is a first attempt to make our approach comparable to other range-intensity correlation algorithms.

\section{Neural Network Gated Depth Estimation}

As described by Sonn et al. in \cite{Sonn2017}, the problem of estimating depth from intensity values of different slices is basically a problem of estimating a function $\mathbf{f}$ that maps these intensity values to depth. 
In \cite{Laurenzis2009}, a simple 5th order ratio-polynomial model was used to approximate depth from a pixel relation value.
In this work, we describe a method to learn such a function with a simple \ac{NN}.

Following \cite{Hornik1989}, a multilayer \ac{NN} with nonlinear activation functions can theoretically approximate any continuous function on a bounded region arbitrarily closely---if the number of neurons is large enough.
A \ac{NN} defines an input-output mapping $\mathbf{f}\left( \boldsymbol{x}; \boldsymbol{\theta} \right)$ by a chain of functions, thus, 
\begin{align}\label{eq:mapping_function}  
\boldsymbol{y} =  \mathbf{f}\left( \boldsymbol{x}; \boldsymbol{\theta} \right)  = \mathbf{f}^{\left(L-1 \right)} \left( \mathbf{f}^{\left(L-2 \right)} \left( \ldots \left( \mathbf{f}^{\left(0 \right)} \left( \boldsymbol{x} \right)  \right) \right) \right)              
\end{align}
where $\boldsymbol{x}$, $\boldsymbol{y}$, $\boldsymbol{\theta}$ and $L$ denote input, output, parameters and depth respectively. 
                
Each function $\mathbf{f}^{\left( l \right)}$ describes a layer that consists of many neurons. 
For every neuron, all of its weighted inputs are added up, a bias is optionally added, and the result is propagated through a nonlinear activation function, e.g., a sigmoid, hyperbolic tangent or a \ac{ReLU}.

By using a training set of known input-output mappings, the optimal parameters $\boldsymbol{\theta}$ can be found that approximate $\mathbf{f}$ best.
Training is done by minimizing a loss function with gradient descent optimization methods and the backpropagation algorithm \cite{Rumelhart1986}.
For a more detailed description of the theory of deep learning, we refer the interested reader to \cite{Goodfellow2016}.

\subsection{Collection of Training Data}

\begin{figure}
\centering
\begin{tikzpicture}
\node (car) at (0,0) {\includegraphics[width=0.6\columnwidth]{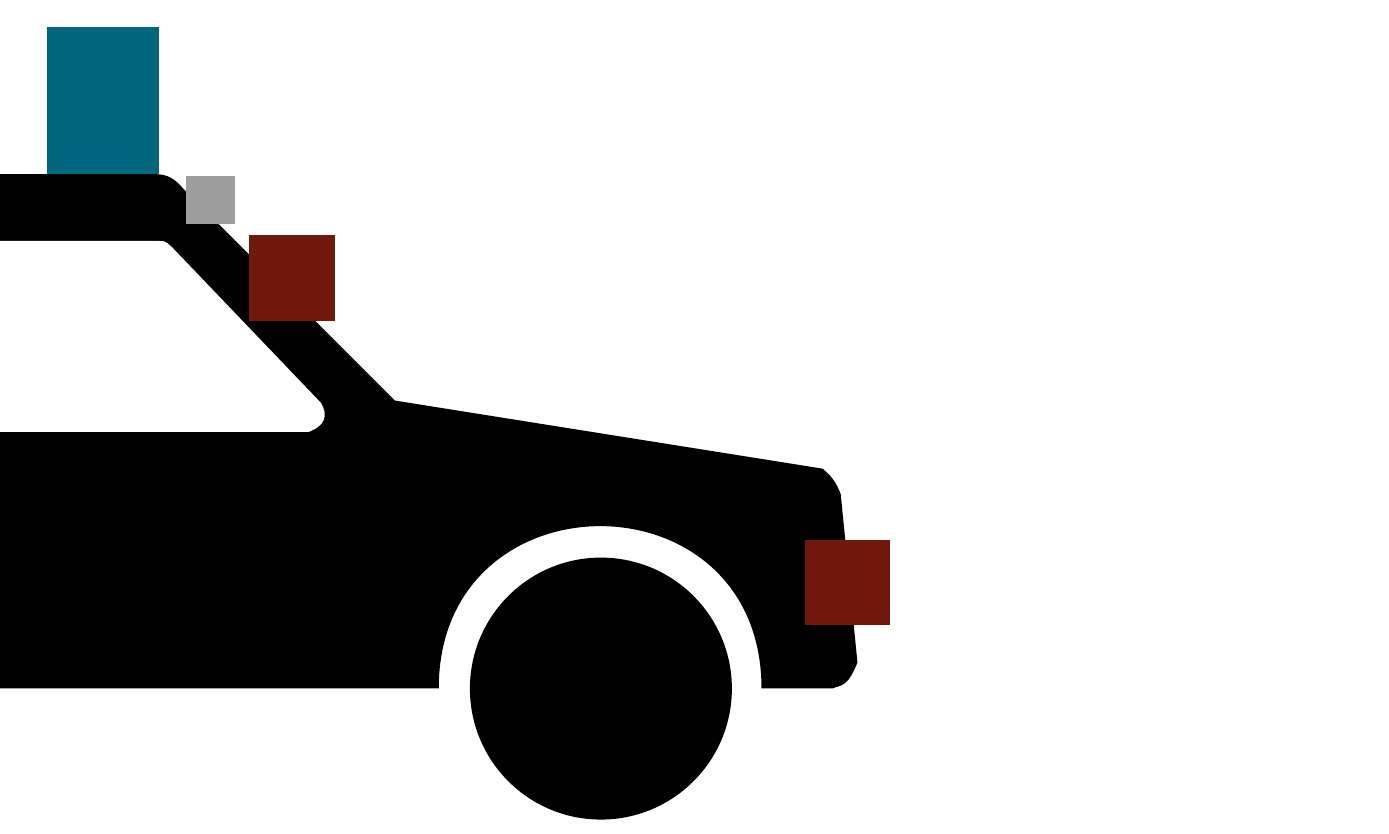}};

\node (pedestrian) at (5,-0.3) {\includegraphics[height=0.1\textheight]{fig/pedestrian.png}};

\node (t) at (6,-1.5) {$r$};

\draw[->, dai_petrol, very thick] (-2.0, 1.2) -- ([yshift=15]pedestrian.west);
\draw[->, dai_ligth_grey40K, very thick] (-1.8, 0.8) -- ([yshift=10]pedestrian.west);

\draw[dai_deepred, very thick] (0.7, -0.65) -- ([yshift=5]pedestrian.west);
\draw[->,dai_deepred, very thick] ([yshift=5]pedestrian.west) -- (-1.6, 0.5);

\draw[->] (-2.6,-1.5) -- (t);

\draw[] (-1.5, -1.4) -- (-1.5, -1.6);
\node at (-1.5,-1.7) {0};

\node[anchor=east] 
at (6,1.5)
{
\begin{tabular}{lll}
 \textcolor{dai_petrol}{\rule{2mm}{2mm}} \footnotesize lidar1 &
 \textcolor{dai_ligth_grey40K}{\rule{2mm}{2mm}} \footnotesize lidar2 &
 \textcolor{dai_deepred}{\rule{2mm}{2mm}} \footnotesize gated camera \\
\end{tabular}
};
\end{tikzpicture}
\caption{Sensor setup: lidar1 - Velodyne HDL64 S3, lidar2 - Velodyne VLP32C, gated camera system - Brightway Vision.}
\label{fig:sensor_setup}
\end{figure}

Data for training the \ac{NN} were recorded by a test vehicle equipped with two \ac{lidar} systems (Velodyne HDL64 S3D and Velodyne VLP32C) and a gated imaging system from Brightway Vision, see Fig.~\ref{fig:sensor_setup}. 
The gated camera offers \unit[8]{bit} images with a resolution of 1280x720 pixels released at a frame rate of \unit[120]{Hz}.
The Velodyne \ac{lidar} systems offer a distance accuracy of \unit[$<$\,2]{cm} and a range of \unit[120]{m} for objects with 80\,\% reflectance.
Intrinsic and extrinsic calibration was performed in order to project \ac{lidar} points into the rectified gated image. 
The different mounting positions of laser illumination and camera can be compensated by setting an additional delay offset.

For the current sensor setup, recording at night is much easier because no additional illumination of the sun has to be considered. 
Nevertheless, active gated imaging at day is not a problem if the laser power is high enough: subtracting a passive image from an active illuminated image yields the typical gated images.
In order to focus on the algorithm and to make things easier, the dataset was recorded at night in Hamburg and Copenhagen. 
In total, we recorded approximately 2 million samples where each consists of three pixel intensities from each slice and its corresponding distance from the \ac{lidar} point cloud.
Since light propagates on the surface of a sphere, we consider the geometric distance $r=\sqrt{x^2 + y^2 + z^2}$ as reference.

We propose to train a \ac{NN} for depth estimation with intensities of three slices as input and distance from our reference system as output.
Nevertheless, some preprocessing of our dataset is required before.

\subsection{Dataset Preprocessing}

\subsubsection{Prefiltering}
 
The raw dataset includes points that are not illuminated or saturated. 
These points do not carry any representative information for the depth reconstruction and have to be filtered out.
\emph{Saturated points} can appear on the image because of the very high reflectance of some objects, e.g. traffic signs. 
All points with the gray pixel values greater than 250 are filtered out and are not used for the further training and evaluation.
\emph{Unilluminated points} are the pixels that do not register any reflected photons back. 
Usually, unilluminated pixels appear in the sky and in the border areas of the camera. 
All samples with a difference between maximum and minimum intensity that is less than 6 are also filtered out.

\subsubsection{Filtering}

Analysis of the training data shows that points exist with the same intensity triples but different distances.
This can be explained by shadowing effects due to different mounting positions of the sensors on the car, projection error due to moving objects and rotating \acp{lidar}, calibration error and \ac{lidar} measurement noise.
In order to improve the quality of our training data, different filtering algorithms are applied.

Our dataset consists of intensity triples and its corresponding distance.
Intensity triples usually appear multiple times with slightly different distances or sometimes in case of errors with completely different distances.
Filtering is based on the mean range for each intensity triple contained in the dataset.
For \texttt{dataset1} and \texttt{dataset2}, we filter out all samples whose range differs more than \unit[1]{m} from the mean range and less than 3 range occurrences. 
Afterwards, the mean is recalculated. 
While \texttt{dataset1} consists only of the recalculated mean value, \texttt{dataset2} includes all filtered samples that may describe the distributions better.

\begin{figure}
	\begin{tikzpicture}
			\begin{axis}
				[
				xlabel=range $r$ / \unit{m},
				ylabel=part of dataset / \,\%,
				xmin=10, xmax=100,
				grid=major,
				width=\columnwidth,
				height=0.25\textheight,
				legend entries={Dataset 1 (9786 points), Dataset 2 (197740 points), Dataset 3 (14223 points), Dataset 4 (902324 points)},
				legend style={
					cells={anchor=west},
					legend pos=north east,
					font=\footnotesize
				},
				]
				
				\addlegendimage{very thick, solid, dai_deepred, mark=*}
				\addlegendimage{very thick, solid, dai_ligth_grey40K, mark=triangle*}
				\addlegendimage{very thick, solid, dai_petrol, mark=*}
				\addlegendimage{very thick, solid, apfelgruen, mark=triangle*}
				
			\addplot+ 	[
						very thick,
						solid, 
						dai_deepred,
						smooth,
						mark=none
						]
			table[x index=0,y index=1,col sep=comma]{data/dataset1.txt};
			
			\addplot+ 	[
						very thick,
						solid, 
						smooth,
						dai_ligth_grey40K,
						mark=none
						]
			table[x index=0,y index=1,col sep=comma]{data/dataset2.txt};
			
			\addplot+ 	[
						very thick,
						solid, 
						smooth,
						dai_petrol,						
						mark=none
						]
			table[x index=0,y index=1,col sep=comma]{data/dataset3.txt};

			\addplot+ 	[
						very thick,
						solid, 
						smooth,
						apfelgruen,						
						mark=none
						]
			table[x index=0,y index=1,col sep=comma]{data/dataset4.txt};

			\end{axis}
		\end{tikzpicture}
	\caption{Number of points in each dataset with respect to distance.}
	\label{fig:number_of_points}
\end{figure}
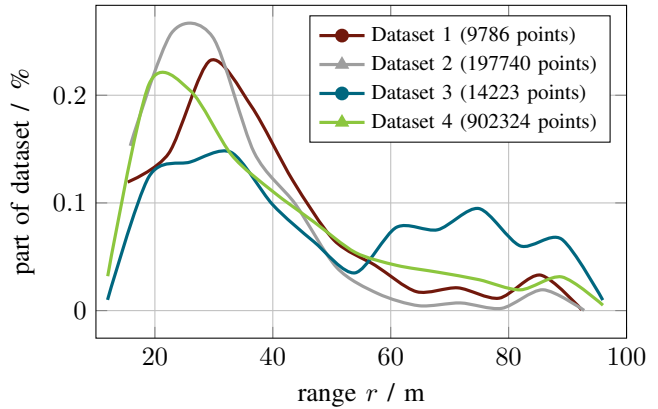

Fig.~\ref{fig:number_of_points} depicts the distribution of the dataset samples.
For distances larger than \unit[60]{m}, there exists only a small amount of training samples.
Therefore, for \texttt{dataset3}, we soften the filter criteria for distances larger than \unit[60]{m} as follows: samples are filtered out if range differs more than \unit[2]{m} and no minimum number of range occurrences is required.

\texttt{dataset4} is the unfiltered dataset and includes all collected points.
It contains a lot of noise and outliers.
We will investigate how good the network can deal with these disturbances.

\subsubsection{Standardization}

In order to make the input values invariant to reflectance and ambient illumination, every sample $S_i = \left( s_{i1}, s_{i2}, s_{i3} \right)$ is standardized by 
\begin{align}
\bar{S}_i = \frac{S_i - \mu_{i}}{\sigma_{i}}
\end{align}
where $\mu_{i} = \frac{1}{3} \sum_{j=1}^3 s_{ij}$ and $\sigma_{i} =\sqrt{\frac{1}{3-1} \sum_{j=1}^{3} \left| s_{ij} - \mu_i \right|^2}$.

\subsection{Network Design}

As there is not yet a standard network for our task, an appropriate network design has to be found. 
The main focus of this work is to present the basic idea and not to find the perfect architecture and structure.
Therefore, we do not spend too much effort and perform just a simple grid search over different network sizes and activation functions.
Grid search means iterating over a finite set of different parameter combinations.
Table~\ref{tab:grid_search} shows the parameter of the grid search.
\begin{table}
\centering
\caption{Grid search parameters.}
\begin{tabular}{ll}

\textbf{Hyper parameters} & \textbf{Values}\\
\hline
Learning rate & 0.1, 0.01, 0.001\\
\hline
Batch size & 4, 8, 16, 32, 64, 128, 256, 512\\
\hline
Network architecture & 5, 10, 20, 40, \\ & 10--5, 20--10, 40--20, \\ & 20--10--5, 40--20--10, \\ & 40--20--10--5 \\
\hline
Activation function & tanh, sigmoid, ReLU\\
\hline
\end{tabular}
\label{tab:grid_search}
\end{table}
We introduce the notation 20--10--5 which describes the architecture of a \ac{NN} employing three hidden layers with 20, 10 and 5 nodes, respectively.

As a loss function, we choose the \ac{MAE}. 
Weights and biases are initialized random uniformly between $\pm 0.05$ and zeros respectively.
We train our network for a maximum number of 100 epochs incorporating an early stopping condition.

In order to assess the grid search, the dataset has to be split into a training set and a validation set.
In our case, we choose 80\,\% for training and 20\,\% for validation.
A test set is recorded separately as explained in the next section.

\begin{figure}
\centering
\begin{subfigure}[t]{\columnwidth}
	\subcaption{Network architecture}
	\vspace{-2mm}
	\begin{tikzpicture}
				\begin{axis}
					[
					ylabel=mean validation loss,			
					ymax=13,
					grid=major,
					legend style={
						cells={anchor=west},
						legend pos=north west,
						font=\scriptsize
					},
					legend entries={dataset1, dataset2, dataset3, dataset4},
					legend columns=2,
					xtick={0, 1, 2, 3, 4, 5, 6, 7, 8, 9},
					xticklabels={$\begin{array}{c} 5 \end{array} $, $\begin{array}{c} 10 \end{array} $, $\begin{array}{c} 20 \end{array} $, $\begin{array}{c} 40 \end{array}$, $\begin{array}{c} 10 \\ 5 \end{array}$, $\begin{array}{c} 20 \\ 10 \end{array}$, $\begin{array}{c} 40 \\ 20 \end{array}$, $\begin{array}{c} 20 \\ 10 \\ 5 \end{array}$, $\begin{array}{c} 40 \\ 20 \\ 10 \end{array}$, $\begin{array}{c} 40 \\ 20 \\ 10 \\ 5 \end{array}$},	
					width=\columnwidth,
					height=0.24\textheight,
					]
					
				\addlegendimage{very thick, solid, dai_deepred, mark=*}
				\addlegendimage{very thick, solid, dai_ligth_grey40K, mark=triangle*}
				\addlegendimage{very thick, solid, dai_petrol, mark=square*}
				\addlegendimage{very thick, solid, apfelgruen, mark=diamond*}
					
				\addplot+ 	[
							very thick,
							solid, 
							dai_deepred,
							mark=*,
							mark size={1.5},
							mark options={dai_deepred}, 
							]
				table[x index=0,y index=1,col sep=comma]{data/grid_search_architecture.data};
				
				\addplot+ 	[
							very thick,
							solid, 
							dai_ligth_grey40K,
							mark=*,
							mark size={1.5},
							mark options={dai_ligth_grey40K}, 
							]
				table[x index=0,y index=2,col sep=comma]{data/grid_search_architecture.data};
				
				\addplot+ 	[
							very thick,
							solid, 
							dai_petrol,						
							mark=square*,
							mark size={1.5},
							mark options={dai_petrol}, 
							]
				table[x index=0,y index=3,col sep=comma]{data/grid_search_architecture.data};
				
				\addplot+ 	[
							very thick,
							solid, 
							apfelgruen,		
							mark=diamond*,
							mark size={1.5},
							mark options={apfelgruen}, 				
							]
				table[x index=0,y index=4,col sep=comma]{data/grid_search_architecture.data};			
									
				\end{axis}
			\end{tikzpicture}
	
	\label{fig:grid_search_architecture}
\end{subfigure}
\begin{subfigure}[t]{\columnwidth}
	\subcaption{Activation function}
	\vspace{-2mm}
	\begin{tikzpicture}
			\begin{axis}
				[
				ylabel=mean validation loss,
				xtick={0, 1, 2},
				xticklabels={tanh, ReLU, sigmoid},
				legend style={
						cells={anchor=west},
						legend pos=north west,
						font=\scriptsize
				},
				legend entries={dataset1, dataset2, dataset3, dataset4},
				legend columns=2,
				ymax = 13,
				grid=major,
				width=\columnwidth,
				height=0.24\textheight,
				]
				
				\addlegendimage{very thick, solid, dai_deepred, mark=*}
				\addlegendimage{very thick, solid, dai_ligth_grey40K, mark=triangle*}
				\addlegendimage{very thick, solid, dai_petrol, mark=square*}
				\addlegendimage{very thick, solid, apfelgruen, mark=diamond*}
				
			\addplot+ 	[
						very thick,
						solid, 
						dai_deepred,
						mark=*,
						mark size={1.5},
						mark options={dai_deepred}, 
						]
			table[x index=0,y index=1,col sep=comma]{data/grid_search_activation.data};
			
			\addplot+ 	[
						very thick,
						solid, 
						dai_ligth_grey40K,
						mark=*,
						mark size={1.5},
						mark options={dai_ligth_grey40K}, 
						]
			table[x index=0,y index=2,col sep=comma]{data/grid_search_activation.data};
			
			\addplot+ 	[
						very thick,
						solid, 
						dai_petrol,							
						mark=square*,
						mark size={1.5},
						mark options={dai_petrol}, 
						]
			table[x index=0,y index=3,col sep=comma]{data/grid_search_activation.data};
			\addplot+ 	[
						very thick,
						solid, 
						apfelgruen,							
						mark=diamond*,
						mark size={1.5},
						mark options={apfelgruen}, 
						]
			table[x index=0,y index=4,col sep=comma]{data/grid_search_activation.data};			
								
			\end{axis}
		\end{tikzpicture}

\label{fig:grid_search_activation}
\end{subfigure}
\caption{Results of grid search.}
\label{fig:grid_search}
\end{figure}
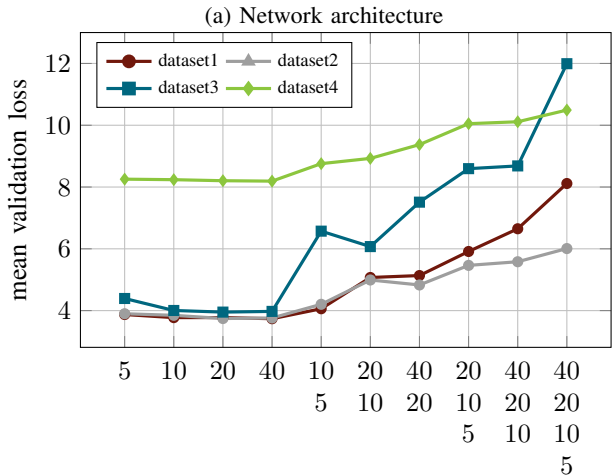
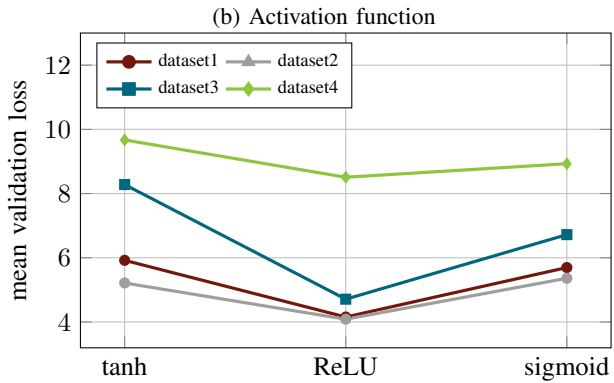

Two main results of the grid search for hyper parameter optimization are shown in Fig.~\ref{fig:grid_search}. 
In order to make the results of different datasets comparable, we search for a single network that performs best for all datasets and not for the best network for each dataset individually.
Fig.~\ref{fig:grid_search_architecture} shows the mean validation loss for different network architectures. 
For each dataset, a simple \ac{NN} with a single hidden layer with 40 nodes seems to be best for our task.
According to Fig.~\ref{fig:grid_search_activation}, the \ac{ReLU} activation function in general performs best. 
In the following evaluation, we use a single hidden layer \ac{NN} with 40 nodes and a \ac{ReLU} activation function.
Training parameters such as learning rate and batch size are adapted individually for each dataset, see Table~\ref{tab:training_parameters}.
\begin{table}
\centering
\caption{Training parameters for each dataset.}
\begin{tabular}{lll}
& \textbf{learning rate} & \textbf{batch size} \\ \hline
dataset1 & 0.1 &  64 \\
dataset2 & 0.01 & 32 \\
dataset3 & 0.01 & 16 \\
dataset4 & 0.01 & 256 \\
\end{tabular}
\label{tab:training_parameters}
\end{table}

\section{Evaluation}

\begin{figure}
\centering
\includegraphics[width=\columnwidth]{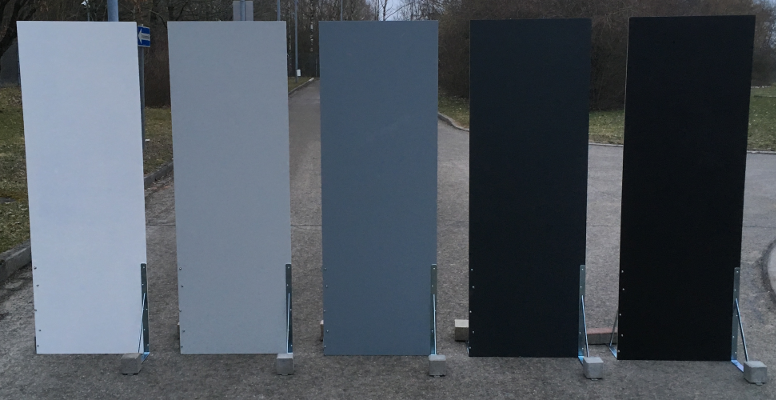}
\caption{Targets for test set. From left to right: white, light gray, middle gray, dark gray, black.}
\label{fig:targets_test_set}
\end{figure}

To evaluate our approach, we set up five targets of \unit[175x60]{cm} with different reflectances, see Fig.~\ref{fig:targets_test_set}.
The \ac{lidar} signals of both systems vanish on the targets at approximately \unit[100]{m}. 
Therefore, we record data for evaluation by slowly driving to the targets, starting from \unit[100]{m} distance and stopping at \unit[10]{m}.
As for the training and validation set, pixel intensity triples and a corresponding \ac{lidar} depth are exported. 
We can only use three targets (middle gray, dark gray, black) for evaluation because the white and light gray target cause too many saturated pixels in all slices and we obtain too few points for evaluation. 
In total, 8103 points are obtained for the black target, 4169 for the dark gray target and 2072 for the middle gray target.

\subsection{Dataset Evaluation}

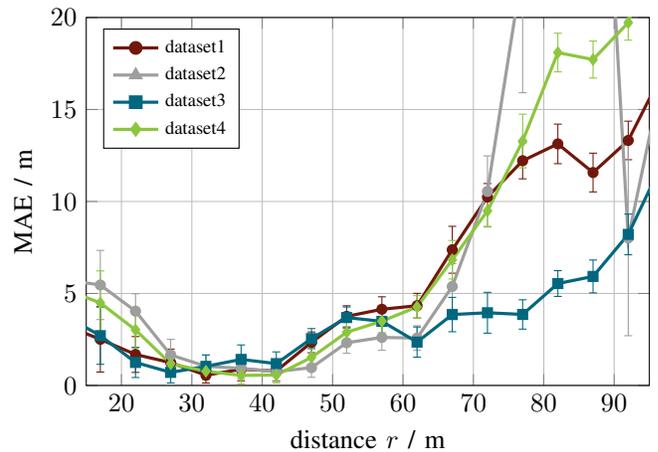
\begin{figure}
	\begin{tikzpicture}
	\begin{axis}[
	xlabel=distance $r$ / \unit{m},
	ylabel=MAE / \unit{m},
	ymin=0,
	ymax=20,
	xmin=15,
	xmax=95,
	grid=major,
	legend style={
		cells={anchor=west},
		legend pos=north west,
		font=\scriptsize
	},
	legend entries={dataset1, dataset2, dataset3, dataset4}, 
	width=1.05\columnwidth,
	height=0.75\columnwidth
	]
	
	\addlegendimage{very thick, solid, dai_deepred, mark=*}
	\addlegendimage{very thick, solid, dai_ligth_grey40K, mark=triangle*}
	\addlegendimage{very thick, solid, dai_petrol, mark=square*}
	\addlegendimage{very thick, solid, apfelgruen, mark=diamond*}
	
	\addplot+ [	very thick, 
				solid, 
				dai_deepred,
				mark=*,
				mark size={1.5},
				mark options={dai_deepred}, 
				error bars/.cd,
				y dir=both,
				y explicit, 
				error mark options={rotate=90,dai_deepred},
	]
	table[col sep=comma, x index=0, y index=1, y error index=2]{data/dataset1/bright_target.txt};
	
	\addplot+ [	very thick, 
					solid, 
					dai_ligth_grey40K,
					mark=*,
					mark size={1.5},
					mark options={dai_ligth_grey40K}, 
					error bars/.cd,
					y dir=both,
					y explicit, 
					error mark options={rotate=90,dai_ligth_grey40K},
		]
		table[col sep=comma, x index=0, y index=1, y error index=2]{data/dataset2/bright_target.txt};
		
	\addplot+ [	very thick, 
					solid, 
					dai_petrol,
					mark=square*,
					mark size={1.5},
					mark options={dai_petrol}, 
					error bars/.cd,
					y dir=both,
					y explicit, 
					error mark options={rotate=90,dai_petrol},
		]
		table[col sep=comma, x index=0, y index=1, y error index=2]{data/dataset3/bright_target.txt};
		
	\addplot+ [	very thick, 
					solid, 
					apfelgruen,
					mark=diamond*,
					mark size={1.5},
					mark options={apfelgruen}, 
					error bars/.cd,
					y dir=both,
					y explicit, 
					error mark options={rotate=90,apfelgruen},
		]
		table[col sep=comma, x index=0, y index=1, y error index=2]{data/dataset4/bright_target.txt};

	\end{axis}
	\end{tikzpicture}
	\caption{Mean absolute error (MAE) for different datasets with respect to distance (middle gray target).}
	\label{fig:dataset_error}
\end{figure}

At first, our different datasets are compared by considering the \ac{MAE} with respect to distance. Fig.~\ref{fig:dataset_error} shows the \ac{MAE} calculated on distance bins of \unit[5]{m}.
The accuracy of our different datasets is quite similar in between \unit[25--45]{m}.
There exist huge differences for small (\unit[$<$\,25]{m}) and large (\unit[$>$\,45]{m}) distances in particular. 
\texttt{dataset2} and \texttt{dataset4}, which contain many noisy points, yield the worst performance for these edge regions. 
This indicates that noisy training data are bad for approximating the function.
It can be clearly seen that the softened filter conditions for large distances in \texttt{dataset3} compared to \texttt{dataset1} help to improve accuracy at greater distances.

For the further evaluation, we use only the \ac{NN} trained with the best performing \texttt{dataset3}.

\subsection{Baseline Comparison}
\begin{figure}[t]
	\begin{tikzpicture}
	
	   \begin{axis}[
	   scale  only  axis,
	   xmin=15,xmax=95,
	   ymax=17,
	   ytick={0,4,8,12,16},
	   axis  y  line*=left,
	   grid=major,
	   legend style={
	   		cells={anchor=west},
	   		at={(0.17,0.86)},
	   		anchor=west,
	   		font=\scriptsize
	   },
	   xlabel=distance $r$ / \unit{m},
	   width=0.73\columnwidth,
	   height=0.73\columnwidth,
	   ylabel=absolute MAE / \unit{m}]
	   
	   \addlegendimage{very thick, solid, dai_deepred, mark=*}
	   \addlegendentry{our approach, absolute}
	   \addlegendimage{very thick, solid, dai_ligth_grey40K, mark=triangle*}
	   \addlegendentry{our approach, relative}
	   \addlegendimage{very thick, densely dotted, dai_deepred, mark=*, mark options={solid}}
	   \addlegendentry{baseline, absolute}
	   \addlegendimage{very thick, densely dotted, dai_ligth_grey40K, mark=triangle*, mark options={solid}}
	   \addlegendentry{baseline, relative}
	   \addplot[
	   		very thick, 
	   	   		solid, 
	   	  		dai_deepred,
	   	 		mark=*,
	   	   		mark size={1.5},
	   			mark options={dai_deepred}, 
	   			error bars/.cd,
	   			y dir=both,
	   			y explicit, 
	   			error mark options={rotate=90,dai_deepred},
	   	] table[col sep=comma, x index=0, y index=1, y error index=2]{data/dataset3/bright_target.txt};
	   \addplot[
	   very thick, 
	   	   	   		densely dotted, 
	   	   	  		dai_deepred,
	   	   	 		mark=*,
	   	   	   		mark size={1.5},
	   	   			mark options={dai_deepred, solid}, 
	   	   			error bars/.cd,
	   	   			y dir=both,
	   	   			y explicit, 
	   	   			error mark options={rotate=90,dai_deepred},
	   	  ] table[col sep=comma, x index=0, y index=1, y error index=2]{data/baseline_bright_target_mae.data};
	   \end{axis}
	   
	   \begin{axis}[
	   scale  only  axis,
	   xmin=15,xmax=95,
	   ymax=0.425,
	   axis  y  line*=right,
	   axis  x  line=none,
	   width=0.73\columnwidth,
	   height=0.73\columnwidth,
	   ylabel=relative MAE]
	   \addplot[
	   		very thick, 
	   		solid, 
	  		dai_ligth_grey40K,
	 		mark=triangle*,
	   		mark size={1.5},
			mark options={dai_ligth_grey40K}, 
			error bars/.cd,
			y dir=both,
			y explicit, 
			error mark options={rotate=90,dai_ligth_grey40K},
	   ]  table[col sep=comma, x index=0, y index=6, y error index=7]{data/bright_target_relativ.txt};
	   \addplot[
	   	   		very thick, 
	   	   		densely dotted, 
	   	  		dai_ligth_grey40K,
	   	 		mark=triangle*,
	   	   		mark size={1.5},
	   			mark options={dai_ligth_grey40K, solid}, 
	   			error bars/.cd,
	   			y dir=both,
	   			y explicit, 
	   			error mark options={rotate=90,dai_ligth_grey40K},
	   	   ]  table[col sep=comma, x index=0, y index=1, y error index=2]{data/baseline_bright_target_mae_rel.data};
	   \end{axis}
		
	\end{tikzpicture}
	\caption{Absolute and relative \acf{MAE} for our approach and our baseline algorithm.}
	\label{fig:baseline_comparison}
\end{figure}
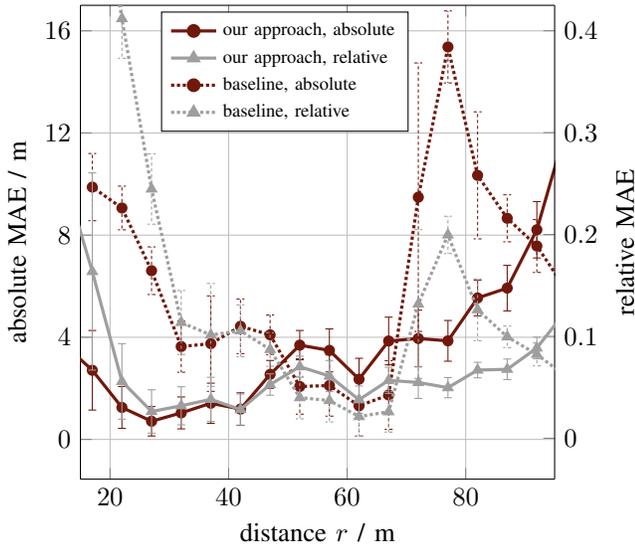

As depicted in Fig.~\ref{fig:baseline_comparison}, our baseline algorithm shows best performance between \unit[50--70]{m}.  
This is reasonable because in this region all three slices overlap.
Our approach trained on \texttt{dataset3} outperforms the baseline algorithm in any region except \unit[50--70]{m}.
As Fig.~\ref{fig:number_of_points} shows, there are quite few training samples for distances larger than \unit[50]{m} which explains the worse performance.
We expect that by collecting more data in this region, the \ac{NN} performance will beat the baseline.
The relative \ac{MAE} in Fig.~\ref{fig:baseline_comparison} shows that for our approach a stable relative accuracy of 5\,\% is achieved in the range between \unit[25--80]{m}.
By design, accuracy is worse for near and far regions because these ranges are covered only by 1-2 slices.
The accuracy of the baseline algorithm varies, probably because in some regions the simulated \acp{RIP} match the real \acp{RIP} better and in other regions worse.

The better performance of our approach suggests that the \ac{NN} is able to capture the real \ac{RIP} shapes of the slices whereas the baseline algorithm is limited to the simulated \acp{RIP} with rectangular shape assumptions. 
While our approach is limited by the range of the \ac{lidar} reference system, the baseline approach can estimate distances for our gated parameters theoretically up to \unit[123]{m}.
Further work should definitely include a baseline approach with more realistic assumptions.

\subsection{Independence Evaluation}

\begin{figure}[t]
	\begin{tikzpicture}
	\begin{axis}[
	xlabel=distance $r$ / \unit{m},
	ylabel=MAE / \unit{m},
	ymin=0,
	ymax=18,
	xmin=15,
	xmax=95,
	grid=major,
	legend style={
		cells={anchor=west},
		at={(0.25,0.81)},
		anchor=west,
		font=\scriptsize
	},
	legend entries={black, dark gray, middle gray, our approach, baseline},
	width=1.03\columnwidth,
	height=0.85\columnwidth
	]
	
	\addlegendimage{very thick, solid, dai_deepred, mark=*}
	\addlegendimage{very thick, solid, dai_ligth_grey40K, mark=triangle*}
	\addlegendimage{very thick, solid, dai_petrol, mark=square*}
	\addlegendimage{very thick, solid, black}
	\addlegendimage{very thick, densely dotted, black}
	
	\addplot+ [	very thick, 
				solid, 
				dai_deepred,
				mark=*,
				mark size={1.5},
				mark options={dai_deepred}, 
				error bars/.cd,
				y dir=both,
				y explicit, 
				error mark options={rotate=90,dai_deepred},
	]
	table[col sep=comma, x index=0, y index=1, y error index=2]{data/dataset3/black_target.txt};
	
	\addplot+ [	very thick, 
					solid, 
					dai_ligth_grey40K,
					mark=triangle*,
					mark size={1.5},
					mark options={dai_ligth_grey40K}, 
					error bars/.cd,
					y dir=both,
					y explicit, 
					error mark options={rotate=90,dai_ligth_grey40K},
		]
		table[col sep=comma, x index=0, y index=1, y error index=2]{data/dataset3/dark_target.txt};
		
	\addplot+ [	very thick, 
					solid, 
					dai_petrol,
					mark=square*,
					mark size={1.5},
					mark options={dai_petrol}, 
					error bars/.cd,
					y dir=both,
					y explicit, 
					error mark options={rotate=90,dai_petrol},
		]
		table[col sep=comma, x index=0, y index=1, y error index=2]{data/dataset3/bright_target.txt};
		
	\addplot+ [	very thick, 
								densely dotted, 
								dai_ligth_grey40K,
								mark=*,
								mark size={1.5},
								mark options={dai_ligth_grey40K, solid}, 
								error bars/.cd,
								y dir=both,
								y explicit, 
								error mark options={rotate=90,dai_ligth_grey40K},
					]
					table[col sep=comma, x index=0, y index=1, y error index=2]{data/baseline_black_target_mae.data};
					
	\addplot+ [	very thick, 
									densely dotted, 
									dai_petrol,
									mark=*,
									mark size={1.5},
									mark options={dai_petrol, solid}, 
									error bars/.cd,
									y dir=both,
									y explicit, 
									error mark options={rotate=90,dai_petrol},
						]
						table[col sep=comma, x index=0, y index=1, y error index=2]{data/baseline_dark_target_mae.data};
						
	\addplot+ [	very thick, 
									densely dotted, 
									dai_deepred,
									mark=square*,
									mark size={1.5},
									mark options={dai_deepred, solid}, 
									error bars/.cd,
									y dir=both,
									y explicit, 
									error mark options={rotate=90,dai_deepred},
						]
						table[col sep=comma, x index=0, y index=1, y error index=2]{data/baseline_bright_target_mae.data};
	
	\end{axis}
	\end{tikzpicture}
	\caption{Mean absolute error (MAE) for different targets (black target, dark gray target, middle gray target).}
	\label{fig:targets_error}
\end{figure}
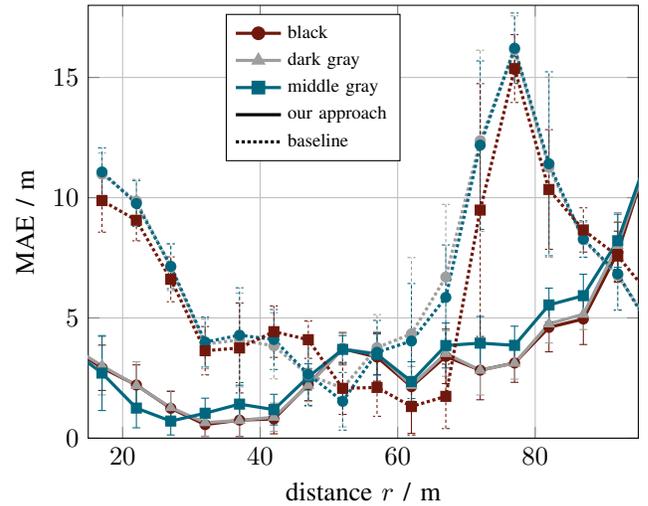

Fig.~\ref{fig:targets_error} depicts the \ac{MAE} and the relative \ac{MAE} for three targets with a different reflectance. 
Both baseline and \ac{NN} algorithm show a similar behavior within the error bars for different reflectances.
This means that our approach is independent of the reflectance of the targets. 
However, if the reflectance of the target is too high, it is possible that the corresponding pixel can be saturated in each slice and therefore no depth information can be captured.

\begin{figure}[t]
	\begin{tikzpicture}
	\begin{axis}[
	xlabel=distance $r$ / \unit{m},
	ylabel=MAE / \unit{m},
	xmin=15,
	xmax=95,
	ymin=0,
	ymax=10,
	grid=major,
	legend style={
		cells={anchor=west},
		legend pos=north west,
		font=\scriptsize
	},
	legend entries={Velodyne HDL64 S3D, Velodyne VLP32},
	width=1.03\columnwidth,
	height=0.55\columnwidth
	]
	
	\addlegendimage{very thick, solid, dai_deepred, mark=*}
	\addlegendimage{very thick, solid, dai_ligth_grey40K, mark=triangle*}
	
	\addplot+ [	very thick, 
				solid, 
				dai_deepred,
				mark=*,
				mark size={1.5},
				mark options={dai_deepred}, 
				error bars/.cd,
				y dir=both,
				y explicit, 
				error mark options={rotate=90,dai_deepred},
	]
	table[col sep=comma, x index=0, y index=1, y error index=2]{data/dataset3/lidar_64.txt};
	
	\addplot+ [	very thick, 
					solid, 
					dai_ligth_grey40K,
					mark=*,
					mark size={1.5},
					mark options={dai_ligth_grey40K}, 
					error bars/.cd,
					y dir=both,
					y explicit, 
					error mark options={rotate=90,dai_ligth_grey40K},
		]
		table[col sep=comma, x index=0, y index=1, y error index=2]{data/dataset3/lidar_32.txt};

	\end{axis}
	\end{tikzpicture}
	\caption{Mean absolute error (MAE) evaluated for different reference measurement systems.}
	\label{fig:lidar_error}
\end{figure}
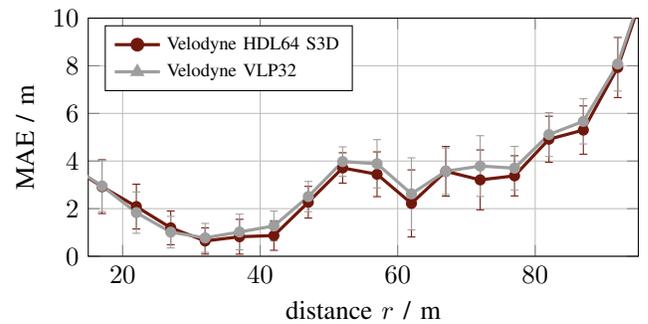

In Fig.~\ref{fig:lidar_error}, one can see the \ac{MAE} taken for two different reference systems. 
Fig.~\ref{fig:lidar_error} clearly shows that our reference measurement system can be considered reliable.

\subsection{Which function was learned?}

Finally, we will take a close look into which function the \ac{NN} has learned. 
Every possible valid intensity triple is fed into our algorithm.
An intensity triple is considered as valid if all values are smaller than 250, the difference between maximum and minimum intensity is larger than 6 and the intensity of the middle slice is not the minimum.
In total, approximately 8 million combinations are evaluated. 
Since the algorithm is trained by \texttt{dataset3} that consists of a subset of all possible combinations (14223 samples, 0.18\,\% of all 8 million combinations), the the \ac{NN} can be considered as interpolator for intensity triples not included in the training dataset.

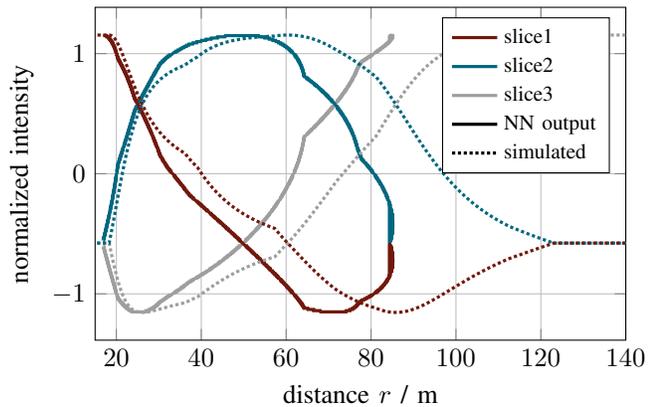
\begin{figure}[t]
	\begin{tikzpicture}
			
			\begin{axis}
								[
								xlabel=distance $r$ / \unit{m},
								ylabel=normalized intensity,
								xmin=15, xmax=140,
								grid=major,
								legend style={
									cells={anchor=west},
									legend pos=north east,
									font=\footnotesize
								},
								legend entries={slice1, slice2, slice3, NN output, simulated},
								width=\columnwidth,
								height=0.25\textheight,
								]
							
							\addlegendimage{very thick, solid, dai_deepred}
							\addlegendimage{very thick, solid, dai_petrol}
							\addlegendimage{very thick, solid, dai_ligth_grey40K}
							\addlegendimage{very thick, solid, black}
							\addlegendimage{very thick, densely dotted, black}
							
							\addplot+ 	[
										solid,
										very thick,
										mark=none,
										dai_ligth_grey40K,
										]
							table[x index=0,y index=1,col sep=comma]{data/slices_230_5step_slice3_cond2.txt};
							
							\addplot+ 	[
										solid,
										very thick,
										mark=none,
										dai_petrol,
										]
							table[x index=0,y index=1,col sep=comma]{data/slices_230_5step_slice2_cond2.txt};
														
							\addplot+ 	[
										solid,
										very thick,
										mark=none,
										dai_deepred,
										]
							table[x index=0,y index=1,col sep=comma]{data/slices_230_5step_slice1_cond2.txt};

											\addplot+ 	[
														very thick,
														densely dotted,
														dai_ligth_grey40K,
														mark=none
														]
											table[x index=0,y index=3,col sep=space]{data/grayvalues_norm.data};
											
											\addplot+ 	[
														very thick,
														densely dotted, 
														dai_petrol,
														mark=none
														]
											table[x index=0,y index=2,col sep=space]{data/grayvalues_norm.data};
											
											\addplot+ 	[
														very thick,
														densely dotted,
														dai_deepred,
														mark=none
														]
											table[x index=0,y index=1,col sep=space]{data/grayvalues_norm.data};

			\end{axis}
		\end{tikzpicture}
	\caption{Learned function illustrated by plotting normalized intensities according to depth. Simulated values are based on the range intensity profiles as depicted in Fig.~\ref{fig:bwv_rip}.}
	\label{fig:learned_function}
\end{figure}

Fig.~\ref{fig:learned_function} illustrates the normalized intensities of each slice according to the estimated depth.
Additionally, the normalized intensities from our simulated \ac{RIP} based on rectangular assumptions are included for reference. 
It can be seen that the normalized simulated \ac{RIP} can be rediscovered in the \ac{NN} output.
Any differences probably come from the fact that in contrast to simulation, \ac{NN} is able to capture the real \ac{RIP}.
Moreover, Fig.~\ref{fig:learned_function} clearly indicates that the \ac{NN} only learns distances in between \unit[18--85]{m}. 
This explains the large errors for the edge regions.

\section{Conclusion}

In this work, we present an approach to estimate distance from active gated image slices by training a \ac{NN}.
This approach is independent from the target reflectance, the reference sensor and the shape of the \acp{RIP}.
We showed, that the \ac{NN} is able to learn the shape of the \acp{RIP} using training data.
Only three slices are required in order to obtain a relative accuracy of 5\,\% for distances between \unit[25--80]{m}.
Thus, a baseline algorithm based on the classical idea of range intensity correlation is outperformed.
Thanks to the resolution of the gated camera (1280x720), a super-resolved depth is obtained and by design perfectly aligned to the gated image.

Since active illumination is required, eye-safety restrictions, interference with other systems and shadowing effects have to be handled. 
In contrast to a \ac{lidar} system it is not possible to adapt the illumination power pixel wise. 
If saturated pixels appear, then no depth restoration is possible.
A gated camera usually has a significantly better range than automotive \ac{lidar} systems.
However, this approach is limited by the range of the \ac{lidar} system.

There is great potential for improving many different aspects of the system.
Pixel-wise distance estimation is done by an extremely small network with a single hidden layer incorporating 40 nodes with a \ac{ReLU} activation function.
If one intend to adapt the \ac{RIP} to certain weather conditions, i.e. small range for foggy weather and maximum range for good weather, fine tuning the network on the fly gives maximum flexibility. 
By using more and better overlapping slices, accuracy will certainly improve.

\section*{Acknowledgment}

The research leading to these results has received funding from the European Union under the H2020 ECSEL Programme as part of the DENSE project, contract number 692449.

\bibliographystyle{IEEEtran}
\bibliography{ref}

\end{document}